\pdfoutput=1
\documentclass[twocolumn,
               showpacs,
               preprintnumbers,
               prd,
               superscriptaddress,
               10pt,
               notitlepage,
               nofootinbib,
               aps]{revtex4-2}
               
\usepackage{graphicx,amssymb,amsmath,amsthm,amsfonts,epsfig,mathtools}

\usepackage[utf8]{inputenc}
\usepackage{graphicx}
\usepackage{float}
\usepackage{dcolumn}
\usepackage{bm}
\usepackage{color}
\usepackage{soul}
\usepackage[dvipsnames]{xcolor}
\usepackage{hyperref}
\hypersetup{colorlinks=true, citecolor=MidnightBlue,linkcolor=CornflowerBlue, urlcolor=CornflowerBlue, linktocpage=true}
\usepackage{xfrac}
\usepackage{siunitx}
\usepackage{empheq}
\usepackage[normalem]{ulem}

\newcommand{\beqn}{\begin{eqnarray}}
\newcommand{\eeqn}{\end{eqnarray}}
\newcommand{\beq}{\begin{equation}}
\newcommand{\eeq}{\end{equation}}

\def\d{\mathrm{d}}

\def\mphi{m_{\phi}}

\def\pt{\tilde{p}}
\def\rt{\tilde{\rho}}

\def\tT{\tilde{T}}

\def\MADM{M_{\textrm{ADM}}}

\newcommand{\Mb}{M_\text{b}}

\begin{document}

\title{Underlying Mechanisms of Phase Transitions in Scalar-Tensor Theories}

\begin{abstract}
Spontaneous scalarization phenomenon in scalar-tensor gravity is known to be a form of phase transition, and it was recently shown that the order of this transition changes depending on the parameters of the theory. There exists a phenomenological description of this result based on Landau theory, but the underlying mechanisms which determine the coefficients of the Landau expansion were unknown. In this study we calculate these coefficients starting from first principles. To this end, we start with an energy functional that describes the nonlinear behavior of the theory, and reduce it to an energy function. This allows us to explain the previously observed, but not well-understood, features of the scalarization phase transition, and enables us to predict which phase transition order will be present for which coupling function or in which regime of the parameter space. The details of the phase transition determine certain astrophysical observables such as signals sourced by transitions from metastable states in first-order scalarization. Thus, predicting these details is an important part of understanding scalarization itself.

\end{abstract}

\author{Murat \"Ozinan}
\email{mozinan21@ku.edu.tr}
\affiliation{Department of Physics, Ko\c{c} University, Rumelifeneri Yolu,
34450 Sar{\i}yer, \.{I}stanbul, T\"{u}rkiye}

\author{K{\i}van\c{c} \.I. \"Unl\"ut\"urk}
\email{kivanc.unluturk@tau.edu.tr}
\affiliation{Department of Electrical and Electronics Engineering, Turkish-German University,\\
34820 Beykoz, \.{I}stanbul, T\"{u}rkiye}

\author{Fethi M. Ramazano\u{g}lu}
\email{framazanoglu@ku.edu.tr}
\affiliation{Department of Physics, Ko\c{c} University, Rumelifeneri Yolu,
34450 Sar{\i}yer, \.{I}stanbul, T\"{u}rkiye}

\date{\today}
\maketitle

\section{Introduction}

Some of the most widely studied ideas in gravitation consider adding a fundamental scalar field to the metric of general relativity (GR)~\cite{Fujii:2003pa}. These \emph{scalar-tensor theories} can lead to a vast array of distinct phenomena depending on how the various coupling terms to the scalar field are chosen. One of the most commonly studied scenarios is spontaneous scalarization, or scalarization in short, where astrophysical objects can grow scalar field clouds depending on their properties such as compactness~\cite{Damour:1993hw,Doneva:2022ewd}. The growth is due to a tachyonic instability which is eventually quenched by nonlinear terms. This phenomenon occurs quite commonly if the coupling terms in the scalar-tensor theory has certain generic features, in other words, it does not require fine tuning. Hence, there are various forms of scalarization called \emph{models,} which had led to a vast literature. Moreover, the final scalarized configurations typically feature large deviations from GR, which has been particularly appealing at the age of gravitational wave astronomy where we can test the dynamical strong-gravity regime~\cite{Will:2006LR,Barack:2018yly,GWTC-4-1, GWTC-4-2, GWTC-4-3}.

The above description of scalarization is reminiscent of a phase transition, which has been noticed since the early days~\cite{Damour:1996ke, Harada:1998ge}, but this fact came to the forefront only recently~\cite{Unluturk:2025zie}. Although first-order or discontinuous transitions were reported before \cite{Harada:1998ge, Novak:1998rk}, scalarization has been studied as a second-order phase transition in most cases. A particularly striking new finding is that we now know it to be first-order in most parts of the theory parameter space for the original exponential coupling considered by Damour and Esposito-Far{\`e}se (DEF)~\cite{Damour:1993hw}. This line of inquiry was also later extended to more general models of scalarization~\cite{Muniz:2025egq, Huang:2025dgc, Herdeiro:2026sur, Staykov:2026ojk}. 

The main difference between a first- and second-order phase transition is continuity. In second-order scalarization, stable scalarized solutions are continuously connected to the unscalarized ones. This is not the case in first-order scalarization in the sense that the least scalarized stable solution already has a finite scalar field, so one needs to ``jump'' to it from the unscalarized branch. First-order scalarization also features metastable solutions which brings the possibility of transitions between scalarized and unscalarized solutions. Overall, a deeper understanding of the phase transition properties of scalarization is not merely a theoretical curiosity, rather, it reveals novel observation channels and tests of gravity~\cite{Kuan:2022oxs,Unluturk:2025zie}.

Recent discoveries on the phase transition nature of scalarization rely on numerical solutions which are time consuming. More crucially, they are also opaque since they do not provide a direct insight to the underlying mechanisms of scalarization. Given a set of theory coupling parameters, we can find all scalarized solutions to arbitrary numerical precision, and learn all there is to know about the phase transition in that specific case. Yet, we have little clue about why things change if the theory coupling parameters are varied or different forms of couplings are considered. This last point is especially important recalling the existence of different models of scalarization. Hence, to obtain a complete picture of the scalarization phase transition with the current approach, one needs to study each model separately, and within each model repeat the same arduous procedure for a large grid on a vast theory parameter space. Such an approach is not only wasteful of time and computational resources, but it also would not answer \emph{why} the phase transition occurs the way it does. For example, we do not know for which models (and within that model for which coupling constants) one observes first- or second-order scalarization. This is a crucial feature of scalarization to learn since it has direct observational consequences as we mentioned.

Our main aim in the current study is addressing this last issue, and obtaining a first-principles understanding of the scalarization phase transition. We develop a methodology that utilizes perturbation theory in order to compute the scalarized solutions of the DEF model via analytical approximations. This ultimately provides a function that expresses the total energy of a scalarized object in terms of the strength of the scalar field. This \emph{energy function} tells us the properties of the phase transition. 

The above explanation is perhaps reminiscent of the Landau theory of phase transitions in condensed matter systems \cite{landau1937theory}, which is not a coincidence. Recall that, in the Landau theory one posits an energy function (or functional), the Landau \emph{ansatz,} without the knowledge of microphysics \cite{Plischke2006, Goldenfeld2018}. The energy function is not derived, it is \emph{designed by hand} to explain the observed qualitative behavior. Such an approach was also successfully applied in the initial analysis of the scalarization phase transition~\cite{Damour:1996ke, Unluturk:2025zie, Muniz:2025egq}. Here we move in the opposite direction, from microphysics to the energy function. We start with the exact nonlinear field equations of the scalar-tensor theory, and derive an energy function whose argument is the scalar field strength. This derivation ensures that the extrema of the energy function give us the correct equilibrium scalar field configurations. We can directly read off the phase transition properties of scalarization from the coefficients of the energy function,  similarly to the phenomenological approach of Landau. The difference from phenomenology is that, we know how the coefficients exactly depend on the theory parameters.

Our new framework enables us to explain some of the surprising observations in the recent literature, for example, the aforementioned observation that first-order scalarization is dominant on the DEF model parameter space. We find that the response of the spacetime and matter to the presence of the scalar field is essential to understand scalarization as a phase transition. This is not an obvious point, since the existence of scalarization can be studied on fixed GR backgrounds, completely disregarding the backreaction.

We use our energy function framework to study the phase transition behavior of scalarization phenomena beyond the original DEF model as well, investigating different forms of coupling and the effect of the scalar mass. This way, we test the power of our approach, and confirm its predictions by fully nonlinear numerical computations. We also provide a map for extending our methodology to other, distinct models of scalarization in the future, opening up new avenues for identifying models of scalarization that are particularly interesting for astrophysics. Examples of first-order scalarization discovered so far occur for neutron stars that are of marginal interest to astrophysics, hence our results are especially important for observational relevance of scalarization.

In Sec.~\ref{sec: scalarization as a phase transition}, we review scalarization as a phase transition and its phenomenological description. In Sec.~\ref{sec: Energy Function Framework for a toy star}, we derive the essential elements of our energy function framework for the simplified example of a constant-density star on a flat background. Sec.~\ref{sec: Energy Function Framework For a Realistic Neutron Star} extends the analysis to fully relativistic neutron stars, and obtain explicit expressions for the energy function.  We explain the previous observations about the scalarization phase transition via first principles in Sec.~\ref{sec: explaining first-order}. In Sec.~\ref{sec: predictions}, we demonstrate how to use our framework to predict the phase transition properties when the DEF model we worked on until that point is modified. We confirm these predictions by comparison to exact numerical results. We end with our conclusions and future directions in Sec.~\ref{sec:conclusions}.

We use geometric units $G=c=1$ throughout the paper. We use the piecewise polytropic HB equation of state (EOS)~\cite{Read:2009yp} for nuclear matter in all our examples, except for Sec.~\ref{sec: eos modification} where we study the effect of different EOS on the phase transition.

\section{Scalarization as a phase transition} \label{sec: scalarization as a phase transition}
We summarize the essential features of spontaneous scalarization and its nature as a phase transition in this section, mainly following \textcite{Unluturk:2025zie}. Readers familiar with this material can skip to the following section.

Quintessential example of spontaneous scalarization arises in the following scalar-tensor theory action~\cite{Damour:1992we,Damour:1993hw}
\begin{align}
\label{eq: stt action}
    S &= \frac{1}{16\pi} \int \d^4x \sqrt{-g} \left(R - 2g^{\mu\nu} \nabla_\mu\phi\nabla_\nu\phi \right)  + S_\text{m}[f_\text{m}, \tilde{g}_{\mu\nu}],
\end{align}
where $\tilde{g}_{\mu\nu} = A(\phi)^2 g_{\mu\nu}$ is the so-called Jordan-frame metric, and $A(\phi)$ is the conformal scaling function . $S_\text{m}$ is the action for the matter fields present, collectively denoted by $f_\text{m}$.\footnote{This original form of scalarization has been mostly ruled out by observations~\cite{Zhao:2022vig}, but this is not the case for massive scalar fields which are still widely studied~\cite{Ramazanoglu:2016kul,Tuna:2022qqr,Demirboga:2023ktt}. However, the scalar mass does not play an essential role in our introductory discussion, and we will study it later, in Sec.~\ref{sec: mphi modification}.} Note that $\phi$ is dimensionless in this action.

Variation of~\eqref{eq: stt action} with respect to $\phi$ yields the scalar-field equation
\begin{equation}
\label{eq: scalar field eqn}
    \square_g\phi = \left[-8\pi A^4 \frac{\d\left(\ln A\right)}{\d\left(\phi^2\right)} \tilde{T} \right]\phi \equiv m_\text{eff}^2 \phi.
\end{equation}
We shall focus on the case originally considered by~\textcite{Damour:1993hw},
\begin{equation}\label{eq: Aphi}
    A(\phi) = e^{\beta\phi^2/2}.
\end{equation}
Then, linearizing Eq.~\eqref{eq: scalar field eqn} around $\phi=0$ and approximating $\tilde{T} \approx -\tilde{\rho}$ for a non-relativistic star, tilde indicating that the quantity is in the Jordan frame, we arrive at
\begin{equation}\label{eq: scalar approximate rho}
    \square_g\phi \approx 4\pi\beta\tilde{\rho}\ \phi.
\end{equation}
We therefore see that for $\beta<0$, the effective mass $m_\text{eff}$ can become imaginary, which leads to a tachyonic instability. In rough terms, the Fourier modes with low wave number $|\mathbf{k}|$ grow exponentially, causing an instability. That is, $\phi=0$ is a solution of the theory (the GR solution), but it is an unstable one. The tachyonic growth can continue until the nonlinear terms come into effect to eventually quench the growth, and the scalar field profile settles at a finite configuration~\cite{Ramazanoglu:2016kul}.

Since its early days, it was known that spontaneous scalarization can be viewed as a phase transition, analogous to spontaneous magnetization, although this phase transition picture was not commonly invoked in the literature. Using Landau theory, \textcite{Damour:1996ke} showed that the scalarization process in their original (DEF) model is formally a second-order phase transition. However, it has been recently shown that this actually holds for a small region of the parameter space of the theory, and the phase transition is first-order in most cases~\cite{Unluturk:2025zie}.

\begin{figure}
    \centering
    \includegraphics[width=\columnwidth]{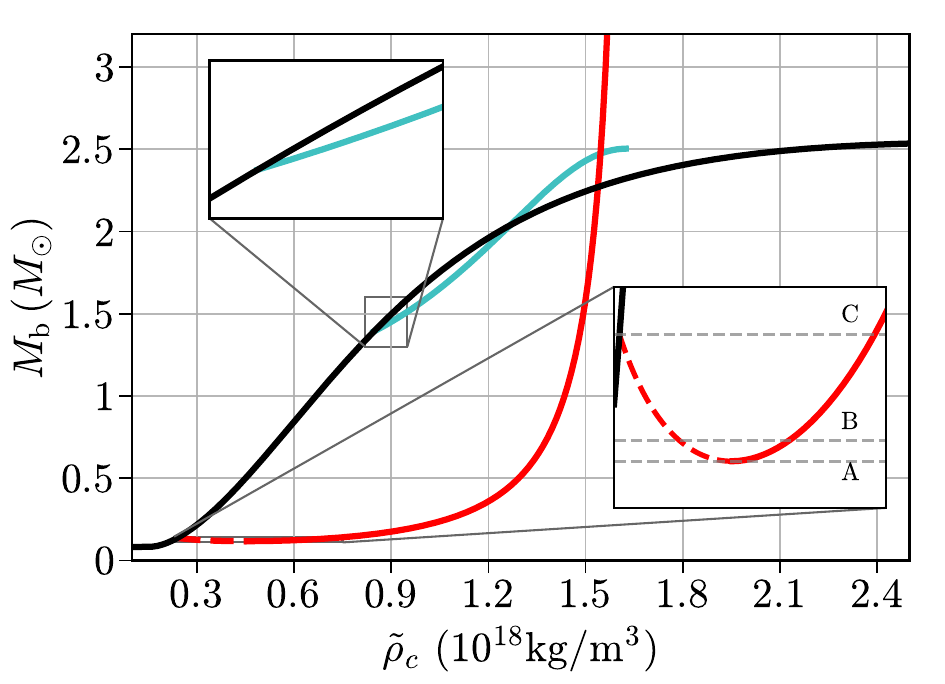}
    \caption{
Baryon mass vs.~the central energy density $\rt_c = \rt(r=0)$ diagrams for GR (black) and scalarized (colored) neutron stars for the original DEF model of scalarization detailed in Eqs.~\eqref{eq: scalar field eqn} and~\eqref{eq: Aphi}. The turquoise curve ($\beta=-5$) shows second-order scalarization, and the red one ($\beta=-30$) first-order scalarization. The dashed portion of the red line represents unstable solutions.} 
    \label{fig: Mb vs rho_c}
\end{figure}

The spherically symmetric, static neutron star solutions in the DEF model for a given value of $\beta$ in Eq.~\eqref{eq: Aphi} form a one-parameter family, where the parameter can be chosen, for instance, as the value of the energy density at the center of the star, $\tilde{\rho}_\text{c} = \tilde{\rho}(r=0)$. Figure~\ref{fig: Mb vs rho_c} shows the baryon masses of these neutron stars as a function of $\tilde{\rho}_\text{c}$ for two different values of $\beta$, as well as the same curve for GR, i.e., for unscalarized stars. 

Let us first look at the ``standard" scalarization picture given by the turquoise curve; we shall comment on the red curve later. The turquoise curve shows the scalarized neutron stars, which branch off from the GR curve at a certain value of the baryon mass $M_\text{b}$. For the specific value of $\beta$ in Figure~\ref{fig: Mb vs rho_c}, there is no scalarized solution below this critical mass $M_\text{crit}$. This means that stars with a baryon mass less than $M_\text{crit}$ do not spontaneously scalarize, whereas stars with baryon mass greater than $M_\text{crit}$ are tachyonically unstable and transition to a scalarized configuration under the slightest perturbation.

This situation is similar to the spontaneous magnetization of a ferromagnet, and can be modeled in the same way using Landau theory. To this end, let us imagine a given matter distribution $\tilde{\rho}(r)$ which can be dressed with a scalar field configuration $\phi(r)$. For simplicity, we suppose that the total energy in the spacetime, i.e. the ADM mass, can be written as an analytic function of the total baryon mass $M_\text{b}$\footnote{There are various mass definitions in general relativity. The baryon mass is the total mass of the matter in a star if we were to separate it into small distant parts. Equivalently, it is the number of baryons in the star multiplied by the average \emph{rest mass} of a baryon. We will further comment on $\Mb$ in Appendix~\ref{appx: numerical}.} and the strength of the scalar field, which we will measure by its value at the center of the star, $\phi_\text{c}$,\footnote{There are other choices as well, such as the scalar charge of the star~\cite{Unluturk:2025zie}.}
\begin{equation}
\label{eq: landau expansion quadratic}
    M_\text{ADM} = M_0(M_\text{b}) + a(M_\text{b})\phi_\text{c}^2 + \frac{1}{2}b(M_\text{b})\phi_\text{c}^4 + \cdots.
\end{equation}
Here, $M_0$ is the ADM mass of an unscalarized star, i.e., a GR star. The coefficients $a$ and $b$ are analytical functions of $M_\text{b}$ and for now we assume $b>0$. We have also used the fact that the theory is symmetric under $\phi\to-\phi$, hence the energy must be an even function of $\phi_\text{c}$.

Now, the baryon mass and the central scalar field of an equilibrium star are not independent, instead, the physically realized $\phi_\text{c}$ for a given $M_\text{b}$ will be the one that minimizes the total energy, Eq.~\eqref{eq: landau expansion quadratic}. Neglecting the higher order terms, we see that for $a>0$, the energy assumes its minimum when $\phi_\text{c}=0$. However, for $a<0$, $\phi_\text{c}=0$ becomes a local maximum, and the two minima of the energy are at $\phi_\text{c} = \pm (-a/b)^{1/2}$. In other words, for $a>0$ there is only one stable equilibrium solution and it is the unscalarized one. For $a<0$, the unscalarized solution $\phi_\text{c}=0$ is an equilibrium, but an unstable one, and the slightest perturbation brings it to either one of the stable equilibria, thus the $\phi\to-\phi$ symmetry gets spontaneously broken.

Keeping in mind that the two solutions $\pm \phi_\text{c}$ correspond to the same point on the $M_\text{b}$ vs. $\tilde{\rho}_\text{c}$ curve in Figure~\ref{fig: Mb vs rho_c}, this is exactly the spontaneous scalarization phenomenon we saw above. We further see that the coefficient $a$ must change sign at the \emph{bifurcation point}, the point where the scalarized solutions branch off from the unscalarized ones in Figure~\ref{fig: Mb vs rho_c}, which corresponds to the baryon mass $M_\text{crit}$. Hence we can write
\begin{subequations}\label{eq: a0 b0}
\begin{align}
    a(M_\text{b}) &= \phantom{b_0 + \ } a_1 \left(M_\text{crit} - M_\text{b}\right) + \cdots,\\
    b(M_\text{b}) &= b_0 + b_1 \left(M_\text{crit} - M_\text{b}\right) + \cdots,
\end{align}
\end{subequations}
with $a_1>0$ and $b_0>0$.

What we have described so far is nothing but the Landau theory for a second-order phase transition, as was originally demonstrated by~\textcite{Damour:1996ke}. Thus, for the value of $\beta=-5$ that corresponds to the turquoise curve in Figure~\ref{fig: Mb vs rho_c}, scalarization is formally a second-order phase transition. The family of stable solutions is sharply divided into two phases: a phase with vanishing scalar field and a phase with nonzero scalar field. The latter family involves solutions with arbitrarily small central scalar fields $\phi_\text{c}$. This means the stable scalarized solutions are connected to the unscalarized ones (at the bifurcation point), which is why a second-order phase transition is also called a continuous phase transition. As we shall soon present, this is not the case for first-order transitions.

As we have mentioned, for more negative values of $\beta$, the situation changes drastically, and we have the red curve in Figure~\ref{fig: Mb vs rho_c}. We can see that it is qualitatively different from the turquoise curve in that the baryon mass decreases with $\tilde{\rho}_\text{c}$ until it reaches a minimum $M_\text{bottom}$, after which it starts to increase. This has the following consequence. The number and nature of the solutions for $M_\text{b} < M_\text{bottom}$ or $M_\text{b} > M_\text{crit}$ is similar to the previous example. Namely, for $M_\text{b} < M_\text{bottom}$, there is only a GR solution, which is stable. For $M_\text{b} > M_\text{crit}$, there is a GR solution and a scalarized one. The GR one is tachyonically unstable, hence goes over to the scalarized solution under a small perturbation, which is the stable one. Between $M_\text{crit}$ and $M_\text{bottom}$, however, we have the novelty that there are three distinct solutions corresponding to the same baryon mass; one unscalarized star and two scalarized ones (up to the $\phi \to -\phi$ symmetry). This raises the question: Which of these would be the physically realized configuration for the given amount of baryonic matter?

We know that the portion of the scalarized branch with decreasing mass is hydrodynamically unstable, hence this part is not physically relevant~\cite{Unluturk:2025zie}. However, the remaining two solutions are, at least locally, stable. The other portion of the scalarized branch with increasing mass is stable, since it is on the other side of a turning point. The GR solution is stable, since there is no hydrodynamical instability in this part of the GR curve and there is no tachyonic instability below the critical mass either. Thus, one of the two locally stable solutions must be the globally stable one, while the other one is metastable. A metastable solution would like to go over to the globally stable one, but cannot do it under small perturbations, since it must overcome an energy barrier. Therefore, both solutions can be potentially observed in nature, depending on the formation history of the star. Also note that, stable scalarized stars cannot have arbitrarily small scalar fields in this case. They are disconnected from the unscalarized solutions, which is manifest in Figure~\ref{fig: Mb vs rho_c}; the stable portion of the red curve is disconnected from the unscalarized (GR) solutions of the black curve.

\begin{figure}
    \centering
    \includegraphics[width=\columnwidth]{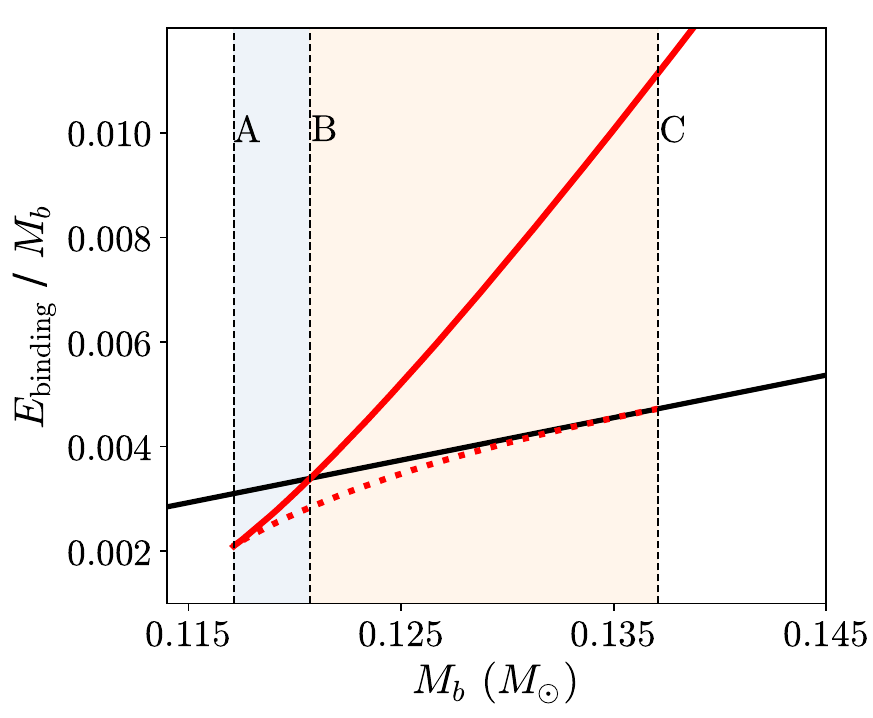}
    \caption{Fractional binding energy $(\Mb- M_{\text{ADM}}) / \Mb$ in the right inset of Figure~\ref{fig: Mb vs rho_c}. Going from lower to higher $\Mb$, metastable scalarized stars begin to exist after line A, in the blue region. In the orange region after line B, it is the GR stars that are metastable, and the scalarized solutions correspond to the global minimum of $\MADM$. To the right of line C, only scalarized solutions are stable.}
    \label{fig: binding energy plot}
\end{figure}

We can determine metastability by checking the total binding energy $E_\text{binding} = M_\text{b} - M_\text{ADM}$ plotted in Figure~\ref{fig: binding energy plot}, which shows that metastability is not exclusive to one branch of solutions. For lower $M_\text{b}$ (blue region between line A and line B) the unscalarized (black) configuration is energetically favored, and the opposite is true for higher $M_\text{b}$ (orange region between lines B and C).

Metastability and discontinuity of solutions are hallmarks of a first-order phase transition. Indeed, the scalarization phenomenon for the red curve is a first-order transition. In order to describe a first-order transition phenomenologically, one needs to consider the next order in the Landau expansion, Eq.~\eqref{eq: landau expansion quadratic}. We therefore expand the ADM mass as
\begin{equation}
\label{eq: landau expansion cubic}
    M_\text{ADM} = M_0(M_\text{b}) + a(M_\text{b})\phi_\text{c}^2 + \frac{1}{2}b(M_\text{b})\phi_\text{c}^4 + \frac{1}{3}c(M_\text{b})\phi_\text{c}^6.
\end{equation}
This time we assume $c>0$ for the overall stability of the system. Previously, we had assumed that $b(M_\text{crit})>0$. In that case, one can check that we can indeed ignore $c$, since it plays no role up to the first order in $(M_\text{b}-M_\text{crit})$, and we have a second-order phase transition when $a$ changes sign. The novelty appears when $b<0$.

\begin{figure}
    \centering
    \includegraphics[width=\columnwidth]{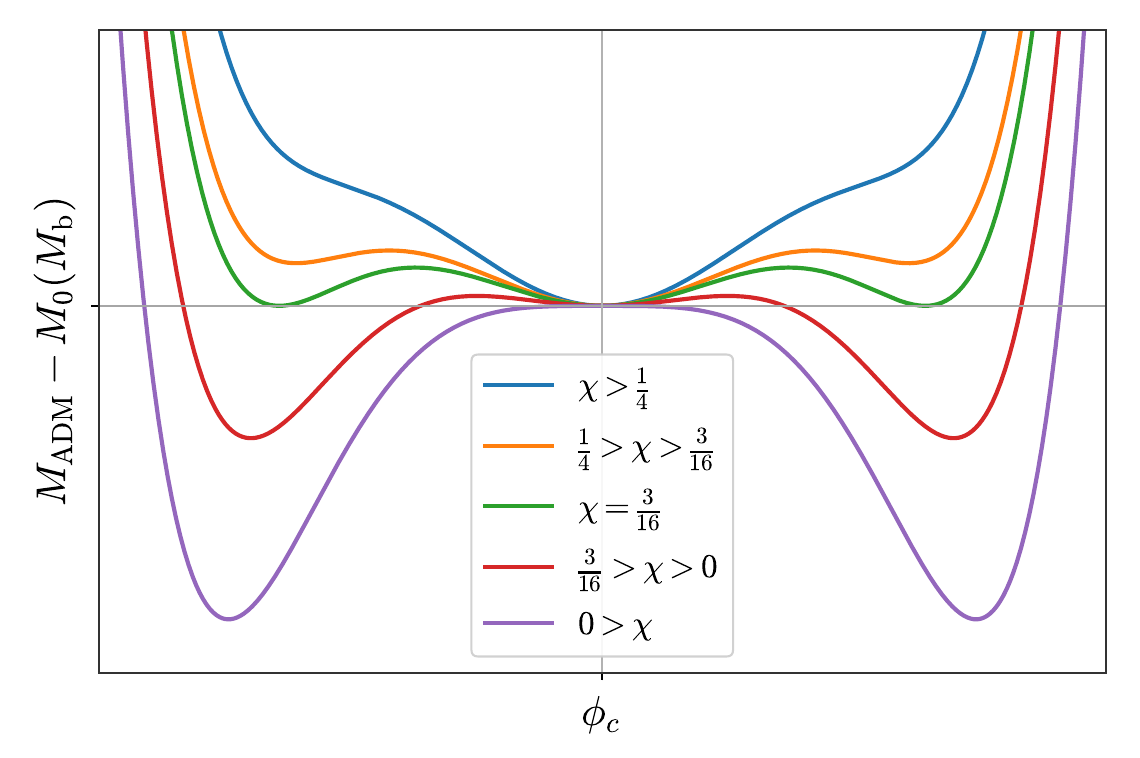}
    \caption{The Landau ansatz in Eq.~\eqref{eq: landau expansion cubic} in the case $b<0$, plotted for different values of the parameter $\chi = ac/b^2$. As explained in the text, this describes a first-order phase transition with metastable equilibria in the intermediate curves.}
    \label{fig: landau expansion first order}
\end{figure}

If we minimize Eq.~\eqref{eq: landau expansion cubic} with respect to $\phi_\text{c}$ in the case of $b<0$, the number and positions of the extrema change depending on the parameter $\chi = ac/b^2$, see Figure~\ref{fig: landau expansion first order}. To
start with, for $\chi>1/4$, the only equilibrium point and the global minimum of Eq.~\eqref{eq: landau expansion cubic} is at $\phi_\text{c}=0$. The only configuration for a given $M_\text{b}$ is therefore stable and unscalarized. We can assume that $\chi$ decreases as $M_\text{b}$ increases, and once $\chi$ drops below $1/4$, local maxima ($-$) and outer local minima ($+$) appear at
\begin{equation}
    \phi_{\text{c}\pm}^2 = \frac{-b}{2c} \left(1 \pm \sqrt{1 - \frac{4ac}{b^2}} \right).
\end{equation}
However, the global minimum is still located at $\phi_\text{c}=0$ if $\chi > 3/16$. This is the case where the unscalarized ($\phi_\text{c}=0$) solution is the globally stable one, and the locally stable scalarized solutions are metastable.

As $M_\text{b}$ increases further, we arrive at $\chi = 3/16$, where the energies at $\phi_\text{c}=0$ and the outer minima become equal. This is the actual point of the first-order phase transition, since it is defined in terms of when the global minimum changes. If $\chi$ becomes less than $3/16$, the scalarized solution at ${\phi_\text{c}}_+$ becomes the globally stable one, and the unscalarized stars become metastable. Note that, at the point of transition, the global minimum jumps \emph{discontinuously} from $\phi_\text{c}=0$ to one of $\phi_\text{c} = \pm {\phi_\text{c}}_+$.

Further increasing $M_\text{b}$ brings us to $a=0$ or $\chi=0$, for which the maxima at $\phi_\text{c} = \pm {\phi_\text{c}}_-$ join the local minimum at $\phi_\text{c}=0$. For negative values of $a$, hence of $\chi$, we have a local maximum at $\phi_\text{c}=0$, and minima at $\phi_\text{c} = \pm {\phi_\text{c}}_+$. This corresponds to having one unstable GR solution and (up to the sign of $\phi_\text{c}$) one stable scalarized solution.

Comparing the cases of first- and second-order scalarization, we quickly see that the main difference lies in the sign of the quartic term $b$ in the Landau energy ansatz. Note that we used these energy formulas in a strictly descriptive manner as one does in phenomenology. Once we have Figure~\ref{fig: Mb vs rho_c}, we already know which order of scalarization occurs for each curve, and then we use the polynomial function that describes the known behavior. However, we would not know what type of scalarization would occur for some other theory parameter, say $\beta=-15$, without computing the scalarized stars in that case.

Perhaps an even more important point is understanding the general trends of the phase transition on the theory parameter space. For example, we have seen that more (less) negative $\beta$ values lead to first-order (second-order) scalarization in the particular case of Figure~\ref{fig: Mb vs rho_c}. This is not an accident, but a general trend. There is no scalarization if $\beta \gtrsim-5$, and scalarization is second-order for $-5 \gtrsim \beta \gtrsim -10$. Afterwards, all $\beta \lesssim -10$ values lead to first-order scalarization. The intervals may change, but the trend is the same even when we deviate from the DEF model in certain ways, for example by adding a scalar mass term: scalarization is always first order once $\beta$ becomes sufficiently negative~\cite{Unluturk:2025zie}. This behavior means $b$ goes to more negative values as $\beta$ goes to more negative values, which can be checked via numerical fits. However, we do not know \emph{why} this behavior occurs, or whether it would change if we considered another model of scalarization, say, one with a different $A(\phi)$. In order to understand these observations, we need to learn how the coefficients of the Landau ansatz such as $b$ depend on the theory parameters such as $\beta$.

Our main task in this study is addressing the above problem. That is, we do not want our energy function as an \emph{ad hoc} ansatz, but rather, we aim to derive formulas like Eq.~\eqref{eq: landau expansion cubic} starting from the action and the equations of motion of the theory. If this is achieved, we learn how $a,b,c$ depend on $\beta$, which would give us the essential phase transition picture without numerically generating an analog of Figure~\ref{fig: Mb vs rho_c} for each separate case of scalarization.

\section{Energy Function Framework for a toy star}\label{sec: Energy Function Framework for a toy star}

\subsection{{Perturbative solutions of scalarization}}
Let us further simplify our case by considering a spacetime that is well approximated by a flat metric, and consider a single spherical harmonic mode of our scalar field
\begin{equation} \label{eq: spherical}
    \phi = \frac{e^{-i\omega t}}{\sqrt{4\pi}} Y_{\ell m}(\theta,\varphi) \frac{u(r)}{r}.
\end{equation}
Linearizing Eq.~\eqref{eq: scalar field eqn} in $\phi$  leads to 
\begin{equation}\label{eq: tachyon first}
    -\frac{\d^2 u}{\d r^2} + \left[
    4\pi \beta \rho + \frac{\ell(\ell+1)}{r^2} \right] u = \omega^2 u.
\end{equation}
This is a functional eigenvalue problem whose solutions provide the modes of the scalar field and their oscillation frequencies $\omega^2$. An instability appears if there exists a mode with $\omega^2 < 0$, since an imaginary frequency would make the corresponding mode grow exponentially rather than oscillate. The $\ell(\ell+1)/r^2$ term is positive definite, so we will only consider the $\ell=0$ mode which is the one most susceptible to scalarization. 

Let us further specify our system to understand this process in finer detail: we consider a toy star of radius $R$, constant density $\rho$ and negligible pressure, while still keeping the flat spacetime approximation. This is not a realistic model for neutron stars, but we will see that it contains most of the essential elements to understand our framework. Let us further concentrate on the very specific case of where the tachyonic instability has just appeared, similar to the bifurcation points of Figure~\ref{fig: Mb vs rho_c}. Since the instability occurs for $\omega^2 < 0$, we will investigate the solution with $\omega=0$.

The scalar field equation~\eqref{eq: tachyon first} for a toy star right at the edge of scalarization reads as
\begin{equation}\label{eq: scalar field main}
     \left[-\frac{\d^2}{\d r^2}-\frac{3C|\beta|}{R^2} H(R-r) \right] u = \omega^2\ u = 0.
\end{equation}
Here, $C=4\pi R^2 \rho/3$ is the compactness of the star, and
\begin{equation}
    H(R-r)=
    \begin{cases} 
      1, &  r < R, \\
      0, & r>R,
   \end{cases}
\end{equation}
is the (shifted and inverted) step function. Recall that $\beta$ has to be negative for scalarization to occur when $\tT \approx -\rho$ as in here, so we will use  $|\beta|$ or $-\beta$ in most of our formulas.\footnote{Realistic neutron stars can also scalarize for positive $\beta$ under exceptional circumstances~\cite{Mendes:2016fby,Mendes:2018qwo}, but we will concentrate on the more common case of negative $\beta$ here.} 

Inside the star, Eq.~\eqref{eq: scalar field main} is solved by
\begin{equation}\label{eq:zero_mode_in}
    u(r) = A \sin\left( \sqrt{\frac{3|\beta|C}{R^2}} r \right), \quad r<R ,
\end{equation}
where the cosine term vanishes due to the boundary condition $u(0)=0$. Outside the star, we simply have $u''=0$, and the only solution that satisfies the boundary condition $\lim_{r \to \infty} u(r)/r=0$\footnote{When the scalar field does not have mass or self-interaction, it can assume nonzero values at infinity, but we will require the scalar field to asymptotically vanish.} is the constant function
\begin{equation}
    u(r) = B\ \ ,\ \ r>R .
\end{equation}
Using the continuity conditions $u(R^-)=u(R^+)$ and $u'(R^-)=u'(R^+)$, we see that there are infinitely many $C$ values that yield a solution, but the smallest compactness that allows for $\omega=0$ is given by\footnote{The eigenfunction that corresponds to the first time we ever encounter a tachyon as we increase the compactness of the toy star is the nodeless one. Even though solutions with nodes are known to appear at higher compactness, scalarized solutions arising from them are unstable in all known cases to the best of our knowledge~\cite{Mendes:2018qwo}.}
\begin{equation}\label{eq: compactness}
     \frac{\pi}{2}=\sqrt{3|\beta|C}=\sqrt{3|\beta|M/R}
     =\sqrt{4\pi |\beta| \rho R^2}.
\end{equation}
For this value we get
\begin{equation}\label{eq:zero_mode}
    u(r)=
    \begin{cases} 
      A \sin\left( \frac{\pi}{2R} r \right), &  r < R, \\
      A, & r>R.
   \end{cases}
\end{equation}
We will call this barely tachyonic solution for $\omega=0$ the \emph{zero mode}.

Recall that not all stars can scalarize, and the criterion $\sqrt{3|\beta|C} = \pi/2$ distinguishes the ones that do. If the compactness is higher, $\sqrt{3|\beta|C}>\pi/2$, the differential operator $\mathcal{D}$ in Eq.~\eqref{eq: scalar field main} has at least one eigenvalue $\omega^2<0$, and scalarization follows. Otherwise, the unscalarized solution is stable to tachyonic growth. This also tells us that stars with compactness $C$ require $|\beta|$ values of $\sim 1/C$ in order to scalarize. Conversely, if $|\beta|$ is of order-of-unity as might be expected from ``naturalness'' arguments, only neutron stars (which have similarly order-of-unity compactness) can scalarize in our universe. This relationship, $|\beta|C \sim 1$, has been known for quite a while~\cite{Ramazanoglu:2016kul}, but our setup will enable us to obtain further results in what follows.

What we have described so far is the tachyonic instability that signals the existence of spontaneous scalarization, however, this description does not tell us anything about the quenching of the growth which leads to the final scalarized star. This latter process requires nonlinear effects whose investigation can quickly become cumbersome, but it is possible with perturbative methods. In this spirit, consider a star which is very slightly above the compactness limit needed for scalarization:
\begin{equation}
     C = C_{\rm crit}+\varepsilon=\frac{\pi^2}{12|\beta|}+\varepsilon ,
\end{equation}
where $C_{\rm crit}$ is the critical compactness from Eq.~\eqref{eq: compactness} and $\varepsilon$ is a small dimensionless parameter for our perturbative expansion. 

If we recover the leading nonlinear correction to Eq.~\eqref{eq: scalar field main} from the series expansion $A(\phi) = 1+ \beta \phi^2 /2 + \cdots$, the differential equation to be solved becomes
\begin{align}\label{eq: scalar field main2}
    \bigg[&-\frac{\d^2}{\d r^2} \\
    &-\left(\frac{\pi^2}{4R^2}+\frac{3|\beta|\varepsilon}{R^2}\right) 
    \left(1 - 2|\beta| \frac{u^2}{r^2} \right) H(R-r) \bigg] u = 0. \nonumber 
\end{align}
Thus, we expect to have an equilibrium scalar configuration with a small amplitude, which we can expand perturbatively as 
\begin{equation}
    u = \varepsilon^\alpha u_1 + \varepsilon^\gamma u_2 + \cdots,
\end{equation}
with $0<\alpha<\gamma$ to be determined. The boundary conditions are the same at all orders: $u_i(0)=0$, $\lim_{r \to \infty} u_i(r)/r=0$ and there is continuity at $r=R.$ 

At the $\mathcal{O}(\varepsilon^\alpha)$ order, we simply have
\begin{equation}\label{eq_order_1}
    \mathcal{D}[u_1]  \equiv \left[-\frac{\d^2}{\d r^2}-\frac{\pi^2}{4R^2} H(R-r)\right] u_1 = 0,
\end{equation}
which has the exact same solution as Eq.~\eqref{eq:zero_mode}, and we have defined the differential operator $\mathcal{D}$ for future reference. $A$ cannot be determined at this leading order of the perturbation theory since the differential equation is a linear and homogeneous one. In other words, we learn the shape of the solution at the leading order, but know nothing about its amplitude. This will later be an important ingredient in building our energy function framework.

The next order in perturbation potentially includes the terms
\begin{equation}\label{eq:gamma order}
    \varepsilon^{\gamma}\ \mathcal{D} [u_2] = \left[\varepsilon^{1+\alpha} \frac{3|\beta|}{R^2}u_1 - \varepsilon^{3\alpha}\frac{\pi^2|\beta|}{2R^2} \frac{u_1^3}{r^2}\right] H(R-r).
\end{equation}
It can be shown that all these terms have to contribute if $A \neq 0$, meaning\footnote{See \textcite{holmes2012introduction} Sec.~6.3 for a similar perturbative solution.}
\begin{align}
    \alpha=\frac{1}{2}, \quad \gamma=\frac{3}{2}.
\end{align}
Thus, the previous (leading) perturbation order was $\mathcal{O}(\varepsilon^{1/2})$, and the current (next-to-leading) order is $\mathcal{O}(\varepsilon^{3/2})$. Hence, the differential equation for the latter, Eq.~\eqref{eq:gamma order}, reads as
\begin{equation}\label{eq_order_3}
    \mathcal{D} [u_2] = \left( \frac{3|\beta|}{R^2}u_1-\frac{\pi^2|\beta|}{2R^2}\frac{u_1^3}{r^2} \right) H(R-r) .
\end{equation}
Note that on the left-hand side, we have the same differential operator as Eq.~\eqref{eq_order_1} from the $\mathcal{O}(\varepsilon^{1/2})$ order of perturbation, but this time it is driven by a source term on the right hand side. We know that the differential operator has a kernel, that is, Eq.~\eqref{eq_order_1} with vanishing source terms has nontrivial solutions. This means $\mathcal{D}$ is not invertible, and a finite solution to Eq.~\eqref{eq_order_3} is only possible if the source is orthogonal to the kernel as a consequence of the Fredholm alternative theorem~\cite{holmes2012introduction}. Thus, we require
\begin{equation}
    0 = \int_0^R \d r\ \left( \frac{3|\beta|}{R^2}u_1-\frac{\pi^2|\beta|}{2R^2}\frac{u_1^3}{r^2} \right) u_1.
\end{equation}
This yields
\begin{equation}
    A^2 = \frac{6}{\pi^2} \frac{\int_0^R \d r\ \sin^2\left( \frac{\pi}{2R} r\right)}{\int_0^R \d r\ \frac{1}{r^2} \sin^4\left( \frac{\pi}{2R} r\right)},
\end{equation}
or, expressed as a dimensionless quantity, 
\begin{equation}\label{eq: solvability}
    k=\frac{A}{R}=\frac{2\sqrt{3}}{\pi \sqrt{\mathcal{S}}} \approx0.618
\end{equation}
is required, where $\mathcal{S} = 2\pi \mathrm{Si}(\pi) - \pi \mathrm{Si}(2\pi) - 4 \approx 3.18$, $\mathrm{Si}(x)= \int_0^x t^{-1}\sin t \,\d t$ being the sine integral function. In other words, the next-to-leading order in perturbation theory does not only provide a further correction to the leading term, but it is necessary to fix the amplitude of the leading term. The exact form of the $\mathcal{O}(\varepsilon^{3/2})$ term $u_2$ is not crucial for our discussion.

\subsection{The energy functional and the energy function}
\label{sec: functional to function}
A second way to approach the problem of scalarization is imposing the perturbation at the level of the action rather than at the level of the field equations. This task is simplified for the toy star since the background is fixed and flat, and the Lagrangian that leads to Eq.~\eqref{eq: scalar field main2} in the stationary case (${\partial^2 u}/{\partial t^2} = 0$) is\footnote{Note that we treat $r$ like a Cartesian coordinate, (i.e., our integration measure is $\d r$, not $r^2 \d r$) since we are already dealing with $u = r \phi$.}
\begin{align} \label{eq: lagrangian}
L &= \int \d r \Bigg[
\frac{1}{2} \left( \frac{\partial u}{\partial t} \right)^2
- \frac{1}{2} \left( \frac{\partial u}{\partial r} \right)^2 \\
& \quad\quad + H(R - r) \left(\frac{\pi^2}{4R^2}+\frac{3|\beta|\varepsilon}{R^2}\right)\left( \frac{1}{2} u^2 - \frac{|\beta|}{2r^2} u^4  \right)\Bigg]. \nonumber
\end{align}
From Eq.~\eqref{eq: lagrangian} we can also obtain the Hamiltonian in the static case
\begin{align} \label{eq: toy_energy_functional}
H = \int_0^R \d r &\Bigg[
\frac{1}{2} \left( \frac{\partial u}{\partial r} \right)^2 \nonumber\\
& - \left(\frac{\pi^2}{4R^2}+\frac{3|\beta|\varepsilon}{R^2}\right) \left( \frac{1}{2} u^2 - \frac{|\beta|}{2r^2} u^4 \right)
\Bigg]. 
\end{align}

The Hamiltonian can directly act as an energy \emph{functional}, it is a function of the scalar field profile $u(r)$ which itself is a function. So, it provides something similar to the Landau ansatz~\eqref{eq: landau expansion cubic}, but not quite so, since our ansatz was a function of real numbers, not function of other functions. We close this gap by utilizing our perturbative solution from the previous subsection. Namely, at the leading order of perturbation theory, we know the exact shape of the solution, but its amplitude is an unknown. So, let us assume $u(r)$ to be of exactly this form, namely, $u(r)=u_1(r)$ which is given in Eq.~\eqref{eq:zero_mode}. This converts our Hamiltonian, a functional of $u(r)$, into a function of energy with the argument $A$. Explicitly
\begin{align}\label{eq: toy_energy_function}
    E(A) &= \frac{1}{16R} A^2 \Bigg[ \pi^2 - 4 \left(\frac{\pi^2}{4R^2}+\frac{3|\beta|\varepsilon}{R^2}\right) R^2 \nonumber\\
    & \quad\quad\quad\quad\quad\quad + 2 \mathcal{S} \left(\frac{\pi^2}{4R^2}+\frac{3|\beta|\varepsilon}{R^2}\right) A^2 |\beta| \Bigg].
\end{align}

To summarize, we have just achieved our stated goal for the simple case of a toy star: we have derived an energy function via first principles whose argument is the strength of scalarization. This is just like the Landau ansatzes we discussed before for the case of fixed baryon mass, as in each curve in Figure~\ref{fig: landau expansion first order}. However, there is the crucial difference that we now know what the coefficients $a,b,c, \dots $ are in terms of the theory parameters, $\beta$ in our case. To be precise, the coefficients depend on $\varepsilon$ and $R$ on top of $\beta$. In this model, the star radius $R$ is a parameter independent of $\rho$, and the knowledge of $R$ is akin to the knowledge of the EOS in the more realistic case. On the other hand, $\varepsilon$ can be seen to be analogous to $\Mb$ in Eq.~\eqref{eq: landau expansion cubic}. In summary, the coefficients $a,b,c,\dots$ are to be seen as functions of $\varepsilon$, and the form of these functions depends on the theory parameter $\beta$ and the EOS, as expected. These points will be more apparent in the coming section where we study realistic stars.

Our proposal is that the scalarized solutions are those that extremize the energy function. One could naturally doubt the effectiveness of our framework; it is a nontrivial statement that among the functions which have the form of the zeroth order solution, the one that minimizes Eq.~\eqref{eq: toy_energy_function} is the correct solution in the limit $\varepsilon\to0$. We test this by comparing its results to what we already know from perturbation theory. A straightforward calculation shows that the nontrivial (positive) minimum of Eq.~\eqref{eq: toy_energy_function} is at
\begin{equation}
A = 2 \sqrt{ \frac{3 \epsilon}{\mathcal{S} (\pi^2 + 12 |\beta| \varepsilon)} } R.
\end{equation}
Expanding in $\varepsilon$, we obtain
\begin{equation}
    A = \frac{2\sqrt{3}}{\pi\sqrt{\mathcal{S}}}R\, \epsilon^{1/2} + \mathcal{O}(\epsilon^{3/2}).
\end{equation}
Hence, to order $\varepsilon^{1/2}$ we get
\begin{equation}
    k = \frac{2\sqrt{3}}{\pi\sqrt{\mathcal{S}}} \approx 0.618,
\end{equation}
which is the same value as the result from the Fredholm alternative theorem in Eq.~\eqref{eq: solvability}.

We demonstrated that our energy function framework provides the correct description of scalarization in the $\varepsilon \to 0$ limit. Naturally, this framework does not lead to accurate results for strongly scalarized solutions. However, recall that many aspects of the phase transition can be surmised by looking at the solutions very close to the bifurcation point, where scalarized solutions (stable or not) continuously connect to unscalarized ones. For example, recall that scalarization is first-order if the curve goes down at this point in Figure~\ref{fig: Mb vs rho_c}. In this sense, the $\varepsilon \to 0$ limit that the energy function framework can handle is already very informative.  

The energy function is not our final destination, we derived it to investigate the properties of the scalarization phase transition. As an initial example, we will investigate the phase transition order for the scalarization of a toy star. We mentioned in Sec.~\ref{sec: scalarization as a phase transition} that the order is controlled by the sign of the quartic term's coefficient $b$ in the energy function. Simply investigating Eq.~\eqref{eq: toy_energy_function} shows that the coefficient of the $A^4$ term is positive for any value of $\beta$ that allows scalarization, which means the scalarization of a toy star in the original DEF model, $A(\phi)=e^{\beta \phi^2/2}$, is always second-order. This might seem at odds with our presentation in Sec.~\ref{sec: scalarization as a phase transition} where phase transitions at different orders appeared for different $\beta$. However, recall that those examples were based on realistic neutron stars, not toy stars, and different physical systems can have distinct phase transition behaviors. Indeed, recent work shows that there are systems where only one type of scalarization occurs~\cite{Huang:2025dgc,Herdeiro:2026sur}, and we will see in Sec.~\ref{sec: explaining first-order} why toy stars can only feature second-order scalarization.

We want to note that our combination of the leading order in the perturbation theory with the energy functional to obtain an energy function also follows the spirit of the phenomenological approach very closely, albeit starting from microphysics and going in the reverse direction. Recall that a Landau energy ansatz such as Eq.~\eqref{eq: landau expansion cubic} considers dressing a star with hypothetical scalar field profiles where the said profiles do not satisfy the field equations in general. The view is that, out of these hypothetical cases, only the ones that extremize the energy ansatz are the true solutions that also satisfy the field equations. Moreover, once we reduce our ansatz to be a function of a single number that represents the strength of the scalar field, we necessarily have assumptions about the shape of the scalar field profile. We followed a very similar logic in our derivation of the energy function when we assumed the leading-order perturbation solution to be the scalar field we have. We exactly know the shape of this solution but its amplitude, which controls the strength of scalarization, is an unknown variable. In a sense, this subsection can be interpreted as providing a concrete setting for the implicit assumptions of the phenomenological approach in Sec.~\ref{sec: scalarization as a phase transition}.

Let us end by summarizing our 3-step algorithm to obtain an energy function.
\begin{enumerate}
    \item Obtain an energy \emph{functional} whose variation leads to the solutions of the theory. Such a function typically exists for the theories of interest to us.
    \item Use the leading order of the perturbation theory, or equivalently the linearized scalar field equations, right at the onset of scalarization (at the bifurcation point) to find the zero mode. Its general shape can be found exactly, but its amplitude will be an unknown.
    \item Assume the scalar field to be in the form of Step~2, and insert it into the energy functional from Step~1. This provides an energy \emph{function} whose argument is the scalar field amplitude from Step~2.
\end{enumerate}

\section{Energy Function Framework For a Realistic Neutron Star} \label{sec: Energy Function Framework For a Realistic Neutron Star}
The main challenge for a relativistic star is the interdependence of the metric, the matter density profile and the scalar field profile, whereas only the scalar field was variable for the toy star. For a spherically symmetric and static metric
\begin{equation} \label{eq: Schwarzschild}
    g_{\mu\nu}\d x^\mu\d x^\mu = - e^{\nu(r)}\d t^2 + \frac{\d r^2}{1- 2\mu(r)/r} + r^2\d \Omega^2,
\end{equation}
the field equations of the scalar-tensor theory lead to a modified version of the Tolman–Oppenheimer–Volkoff (TOV) equations
\begin{subequations}\label{eq: tov full}
\begin{align} 
    \mu' &= 4\pi r^2 A^4 \rt + \frac{1}{2}r(r-2\mu) \psi^2, \\
    \nu' &= r\psi^2 + \frac{r^2}{(r-2\mu)}\left[8\pi A^4\pt+2\frac{\mu}{r^3} \right], \\
    \phi'&= \psi,\\
    \psi' (r-2\mu) &= 4\pi r A^4 \left[\alpha(\rt-3\pt)+r\psi(\rt-\pt)\right] \label{eq: tov_psi}  \nonumber\\
    & \qquad -2\psi(1-\mu/r),  \\
    \pt' &= -(\rt+\pt)\left( \nu'/2+\alpha \psi \right).  
\end{align}
\end{subequations}
Here, $'$ denotes a derivative with respect to $r$, and $\alpha = {\d\ln A}/{\d\phi} = \beta \phi$. We assume the nuclear matter to be a perfect fluid, and the TOV system is closed by the EOS $\rt(\pt)$.  Note that the $\nu'$ equation is auxiliary. $\nu$ does not appear anywhere on the right hand side, and we can simply insert the $\nu'$ equation into the $\pt'$ equation to remove it from the system. Nevertheless, we will keep it in our formulas to follow the literature and to have a shorter expression for $\pt'$.

The first step of our algorithm to obtain an energy function is identifying an energy functional. We already know that the total energy of the spacetimes we are interested in is given by the Arnowitt-Deser-Misner (ADM) mass which is equal to $\mu(r=\infty)$:
\begin{equation}\label{eq: ADM mass}
    \MADM = \int_0^\infty \d r \left[ 4\pi r^2 A^4 \rt + \frac{1}{2}r(r-2\mu) \psi^2 \right] .
\end{equation}
At this point, let us also give the formula for the baryon mass,
\begin{align}\label{eq: baryon mass}
    M_\text{b} = \int_0^{\infty} \ 4\pi r^2 \tilde{\rho}_\text{r} A(\phi)^3 \left(1-\frac{2\mu}{r} \right)^{-1/2}  \d r ,
\end{align}
where $\tilde{\rho}_\text{r}$ is the rest mass density.\footnote{Although the integral in Eq.~\eqref{eq: baryon mass} runs up to $r=\infty$, the mass density $\rt_\text{r}$ vanishes outside the star. Hence, in practice, one integrates only up to the radius $R$ of the star.} The details of the relationships between the pressure $\pt$, total energy density $\rt$ and the rest-mass density $\tilde{\rho}_\text{r}$ are not essential aside from two facts. First, these three are all increasing functions of each other. Second, the \emph{stiffness} of the nuclear matter, the rate of increase of pressure as density changes, also increases with density: $\d\pt/\d\rt$, hence the speed of sound, increases with $\rt$. We will revisit these when we try to explain the dominance of first-order scalarization with increasing $|\beta|$ via first principles. We employ the HB EOS~\cite{Read:2009yp} in this section, but the essential features we mention are typical for realistic EOS.

The second step of the algorithm is finding the zero mode at the bifurcation point. We again switch to the variable $u(r) = r \phi(r)$, and use the tortoise coordinates $\d r_* = e^{\nu_0/2} (1 - 2\mu_0/r)^{-1/2}\, \d r $, for which the linearized field equation is of the Schr\"odinger form
\begin{equation} \label{eq:schrodinger}
-\frac{\d^{2}}{\d r_{*}^{2}} u
+ V(r_*) u = 0,
\end{equation}
where 
\begin{equation}\label{eq: zero mode potential}
    V(r_*) =e^{\nu_0} \left[\frac{2\mu_0}{r^{3}}
+ 4\pi (\pt_0 - \rt_0) + 4\pi \beta (\rt_0 - 3\pt_0) \right].
\end{equation}
This form makes a numerical solution easier, whose details are in Appendix~\ref{appx: numerical}. The zero subscripts indicate that these are the values for a GR star at the bifurcation point.

\begin{figure}
    \centering
    \includegraphics[width=\columnwidth]{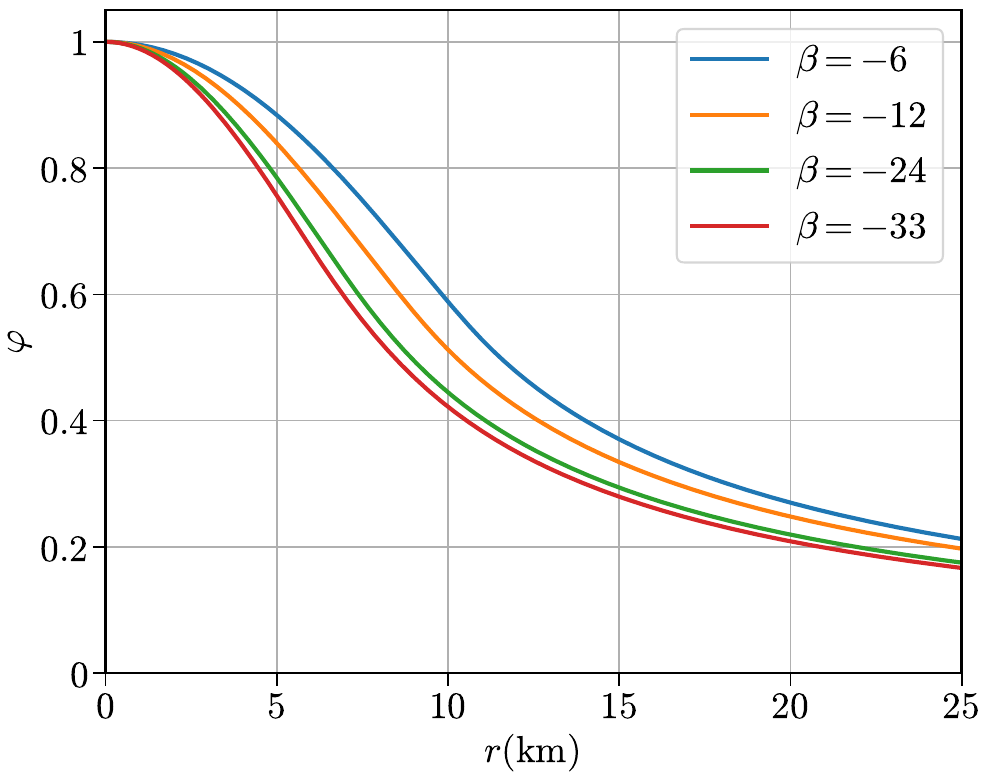}
    \caption{The zero mode $\varphi$ from Eq.~\eqref{eq: phic} for various $\beta$. The functional profiles are relatively similar inside the stars, whose radii are in the 11-13 km range.}
    \label{fig: tachyon}
\end{figure}
We obtain the zero-mode solution numerically for a realistic star, unlike the fully analytic calculation for a toy star. However, it is a relatively simple calculation and the essential features are the same as the toy star. That is, this tachyonic mode is still the solution to a linear equation, hence we can find its exact shape, but would not know its amplitude:
\begin{equation}\label{eq: phic}
    \phi(r) = \phi_\text{c} \varphi(r),
\end{equation}
where $\varphi(r)$ specifies the shape of the zero mode with the specification of $\varphi(0)=1$, and consequently $\psi(r) = \phi_\text{c} \varphi'(r)$. Furthermore, we are still considering the nodeless ground state solution, so we have a rough idea about what $\varphi(r)$ graphically looks like. $\phi_\text{c}=\phi(0)$ is the central value of the scalar field, similar to $A$ in Sec.~\ref{sec: functional to function}. To repeat ourselves, we are not trying to find a perturbative series for the scalar field. We assume Eq.~\eqref{eq: phic} represents the scalar adequately, and we will try to find all other relevant quantities in terms of its amplitude $\phi_\text{c}$, just as in the case of the toy star. We plot $\varphi(r)$ for different values of $\beta$ in Figure~\ref{fig: tachyon} to provide a general idea about the functional form. It is nodeless and monotonically decreasing in all cases, decreasing to roughly half of its central value at the stellar surface. Note that the $\beta$ dependence is quite weak, at least inside the star. 

In the third and last step of our algorithm, we insert the above form of the scalar field into the energy functional~\eqref{eq: ADM mass} to obtain an energy function. In other words, for a given baryon mass $\Mb$, we want to obtain the formula\footnote{The next term in the expansion, $c$, needs to be positive for global stability just like before. However, we only need the first two terms to learn the main properties of the phase transition, hence we will only expand our perturbative calculation to this order.}
\begin{equation}\label{eq: ADM mass expansion}
    \MADM = M_0 + a\,\phi_\text{c}^2 + \frac{1}{2}b\,\phi_\text{c}^4 + \cdots.
\end{equation}
Note that $\MADM$ is a function of both the baryon mass $\Mb$ and the scalar field strength as in Eq.~\eqref{eq: landau expansion cubic}, but we suppressed the $\Mb$ dependence of the coefficients, since we are only interested in the scalar field dependence of the energy for a fixed $\Mb$ value. In other words, we want to learn what happens to the total energy when a neutron star with a fixed number of baryons (no change in total matter content) is dressed with scalar field profiles of different amplitudes. Furthermore, we will use this expansion only for solutions very close to the bifurcation point where $\phi_\text{c}$ is tiny, which is enough to obtain the order of scalarization. Hence, we expect Eq.~\eqref{eq: ADM mass expansion} to be a reasonably accurate approximation. We specifically mention this since it is known that an accurate description far away from $\phi_\text{c}=0$ requires a higher-order polynomial, especially for first-order transitions~\cite{Muniz:2025egq}.

Obtaining the explicit form of Eq.~\eqref{eq: ADM mass expansion} presents a new technical challenge: $\phi$ is not the only variable for a realistic star. The interaction of the scalar field with, for example, the matter distribution $\rt(r)$ means that it will depend on $\phi_\text{c}$ as well. To reflect this interaction, we express all our field variables as series in $\phi_\text{c}$
\begin{subequations}
\begin{align}
    \mu(r) &= \mu_0(r) + |\beta| \phi_\text{c}^2\,\mu_2(r) + |\beta|^2 \phi_\text{c}^4\,\mu_4(r) + \cdots,\\
    \nu(r) &= \nu_0(r) + |\beta| \phi_\text{c}^2\,\nu_2(r) + |\beta|^2 \phi_\text{c}^4\,\nu_4(r) +\cdots,\\
     \pt(r) &= \pt_0(r) + |\beta| \phi_\text{c}^2\, \pt_2(r) +  |\beta|^2\phi_\text{c}^4\, \pt_4(r)  + \cdots,\\ 
     \rt(r) &= \rt_{0}(r) + |\beta| \phi_\text{c}^2\, \rt_{2}(r) +|\beta|^2 \phi_\text{c}^4\,\rt_{4}(r) + \cdots.
\end{align}
\label{eq: tov perturbative expansion}
\end{subequations}
$\pt_{2,4}$ and $\rt_{2,4}$ terms represent the response of the matter profile to the scalar field, which are coupled through the $A(\phi)$ term of the theory. Similarly, $\mu_{2,4}$ and $\nu_{2,4}$ terms are related to the response of the spacetime, often called the backreaction. Note that $\phi$ couples to matter only through the combination $\beta\phi^2$. Hence, we expect the response terms at the $\phi_\text{c}^{2n}$ order, such as $\rt_{2n}$ to naturally increase by a factor of $|\beta|^n$. We make this natural $\beta$ dependence explicit in Eq.~\eqref{eq: tov perturbative expansion} and isolate it from other factors that affect the magnitude of these terms, which we will discuss later.

The response terms play an essential role in the phase transition properties of scalarization as we will soon see. We will not need the explicit form of the systems of differential equations that arise from inserting the perturbative expansion~\eqref{eq: tov perturbative expansion} into the TOV equation~\eqref{eq: tov full} for our general discussion here. They can be found in Appendix~\ref{appx: response contribution}.

Inserting the series expansions of $\mu, \pt$ into the $\MADM$ formula~\eqref{eq: ADM mass} and combining the $\phi_\text{c}^n$ terms at the same power, we can finally obtain the series expansion coefficients in Eq.~\eqref{eq: ADM mass expansion}. The coefficient of the $\phi_\text{c}^2$ term is
\begin{equation}\label{eq: a integral}
    a_0 = |\beta|\int_0^{\infty} \d r\ 4\pi r^2 \left[ -2\varphi^2 \rt_0
    + \rt_2
    +\frac{\varphi'^2}{8\pi|\beta|}\left(1-\frac{2\mu_0}{r}\right)
    \right],
\end{equation}
the subscript $0$ again indicating that we are calculating this value at the bifurcation point. Recall that $a=0$ is the point where the GR solutions start to become tachyonically unstable, that is, the bifurcation point. Hence, the $a_0$ integral should vanish within the numerical and approximation precisions of our framework. Thus, this formula does not directly inform us about the phase transition, but we can  and did use $a_0=0$ as a sanity check for our framework and numerics.

The essence of the phase transition information is in the $\phi_\text{c}^4$ term
%
\begin{align}\label{eq: b integral}
    \frac{1}{4\pi |\beta|^2}\ \frac{b_0}{2} 
    &= I_0 +I_2 +I_4 +I_\mu \nonumber\\
    &= \int_0^{\infty} \d r\, r^2 \left(2 \varphi^4 \rt_0 \right) \nonumber \\
    & +\int_0^{\infty} \d r\, r^2 \left(  
    - 2 \varphi^2 \rt_2 \right) \nonumber \\
    & +\int_0^{\infty} \d r\, r^2  \rt_4 \nonumber \\
    & +\int_0^{\infty} \d r\, r^2 \left(- \frac{\varphi'^2 \mu_2}{4\pi|\beta|r} \right),
\end{align}
which is central to all our subsequent discussion. $I_{0,2,4}$ and $I_\mu$ correspond to the integrals on separate lines that contain the terms $\rt_{0,2,4}$ and $\mu_2$, respectively. Recall that we showed in Sec.~\ref{sec: scalarization as a phase transition} that it determines the order of the phase transition. Scalarization is first-order (second-order) when $b_0$ is negative (positive). Hence, the sign of $b_0$ will reveal whether features unique to first-order scalarization, such as a discontinuous transition and the possibility of locally stable scalarized and unscalarized configurations at the same $\Mb$, are present or not. Eq.~\eqref{eq: b integral} might look complicated at first sight, but each term, including the shape of the tachyonic mode $\varphi$ from Eq.~\eqref{eq: phic} will have been (numerically) computed at the end of our perturbative approach, so it is a relatively straightforward exercise to obtain it once our machinery is set up.

\begin{figure}
    \centering
    \includegraphics[width=\columnwidth]{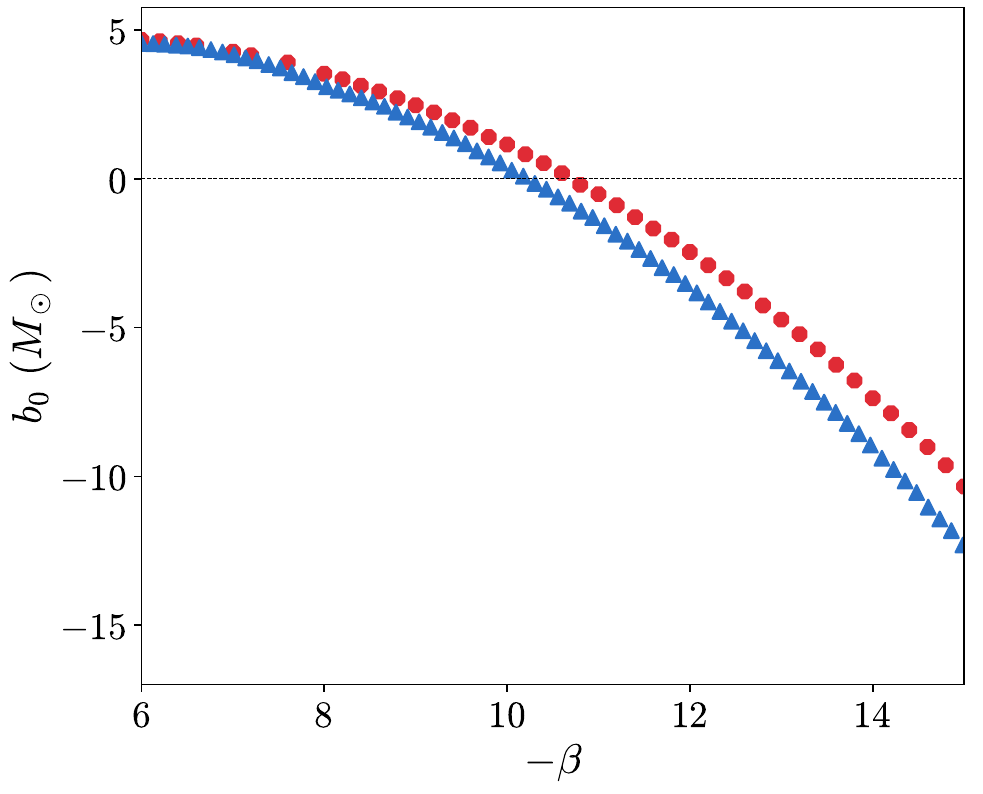}
    \caption{Comparison of the $b_0$ values obtained by the new energy function framework via Eq.~\eqref{eq: b integral} (blue triangles), and numerical computations via Eq.~\eqref{eq: fit b_0} (red circles). The energy function captures all the qualitative features of the $b_0(\beta)$ curve, as well as providing quantitative agreement with reasonable error, which is possibly a result of our perturbation approximation becoming less accurate as $|\beta|$ increases.}
    \label{fig: b0 vs beta}
\end{figure}
Before we start interpreting the terms of Eq.~\eqref{eq: b integral} and use it to gain physical insight into scalarization as a phase transition, let us first check if it agrees with what we already know. Once we have all the scalarized and unscalarized neutron star solutions for various theory parameters (only $\beta$ in our simple scalarization model), we can numerically obtain the dependence of $b_0$ on $\beta$. This is possible thanks to the fact that scalarized solutions sufficiently close to the bifurcation point satisfy
\begin{equation}
\label{eq: fit b_0}
    2(M_0 -M_\text{ADM}) \approx b_0\, \phi_\text{c}^4,
\end{equation}
and all terms aside from $b_0$ can be computed when we numerically construct scalarized stars by integrating the exact nonlinear TOV system~\eqref{eq: tov full}. So, a simple fit provides $b_0(\beta)$~\cite{Unluturk:2025zie}.

Figure~\ref{fig: b0 vs beta} compares the results from such a fit to what the integral formula for $b_0$ in Eq.~\eqref{eq: b integral} provides, validating the latter. The agreement is not perfect, but quite close.\footnote{One factor for the difference might be the fact that our perturbative parameter $\beta \phi_\text{c}^2$ becomes bigger with $\beta$ since we cannot go to arbitrarily small $\phi_\text{c}$ in numerical computations. This means the approximation errors naturally increase with $|\beta|$.} More essentially for our purposes, Eq.~\eqref{eq: b integral} leads to a decreasing $b_0(\beta)$ as $\beta$ becomes more negative. Recall from Sec.~\ref{sec: scalarization as a phase transition} that first-order scalarization occurs in the case of $b_0<0$, and first-order scalarization is also known to be the only type of scalarization we observe when $\beta$ becomes sufficiently negative. Hence, Figure~\ref{fig: b0 vs beta} indeed shows that Eq.~\eqref{eq: b integral} explains this observation. The dominance of first-order scalarization on the theory parameter space, the one-dimensional $\beta$ line in this case, could be a prediction of our energy function framework, if it had not been already established through extensive numerical computation~\cite{Unluturk:2025zie}.

\section{Understanding First-order Scalarization via first principles} \label{sec: explaining first-order}
Let us start our physical interpretation of Eq.~\eqref{eq: b integral} by a comparison of the realistic and toy stars, which provides one of the main lessons of this work: backreaction of the spacetime and the response of stellar matter to the scalar field are essential to understand scalarization as a phase transition. Recall that the toy star has no such response, its matter profile and spacetime are fixed. In the language of Eq.~\eqref{eq: b integral}, this means that the three terms containing $\rt_2, \rt_4, \mu_2$ are absent for the toy star, since they represent the response to the scalar field. Hence, the only nonzero term for the toy star is $2\varphi^4 \rt_0$, which is arguably the easiest one to interpret: it is always manifestly positive, making $b_0$ positive as well. This means we only observe second-order scalarization for a toy star as inferred in Sec.~\ref{sec: functional to function}. 

We already saw that realistic stars feature both second- and first-order scalarization depending on whether $b_0$ is positive or negative, respectively. This means that the cases with first-order scalarization are the ones where the response of matter and spacetime to the scalar field dominates Eq.~\eqref{eq: b integral}, and the overall contribution of these terms should be negative to balance out the term $2 \varphi^4 \rt_0$. We also mentioned the more specific observation that first-order scalarization is not only possible for realistic stars, but it is also dominant in the sense that we observe only first-order scalarization once $\beta$ becomes sufficiently negative, e.g. for $\beta \lesssim -10$ in Fig.~\ref{fig: b0 vs beta}. So, we now want to use the formula~\eqref{eq: b integral} for $b_0$ provided by our energy function framework to explain these two main points:
\begin{enumerate}
    \item Why do we see first-order scalarization for realistic stars at all?
    \item Why does first-order scalarization become the default outcome when $\beta$ becomes sufficiently negative?
\end{enumerate}
We already see these facts in Figure~\ref{fig: b0 vs beta}. What we are after now is gaining an insight to \emph{how} $b_0(\beta)$ arises from a basic physical understanding of $\rt_{0,2,4}$ and $\mu_2$.

To begin with, note that there is an overall factor of $|\beta|^2$ in Eq.~\eqref{eq: b integral}, hence whatever value obtained from the integral will get enhanced as $\beta$ becomes more negative. This does not tell us anything about the sign of $b_0$, but it partially explains the overall increase in magnitude of $b_0$ as seen in Figure~\ref{fig: b0 vs beta}. Note that this $|\beta|^2$ factor is a reflection of $|\beta|$ measuring the strength of the coupling between matter and the scalar field as we discussed in relation to Eq.~\eqref{eq: tov perturbative expansion}. Hence, the overall $|\beta|^2$ factor isolates the \emph{direct} effect of the coupling strength, so that we can separate the indirect effects in the integrands of Eq.~\eqref{eq: b integral}.

\begin{figure}
    \centering
    \includegraphics[width=\columnwidth]{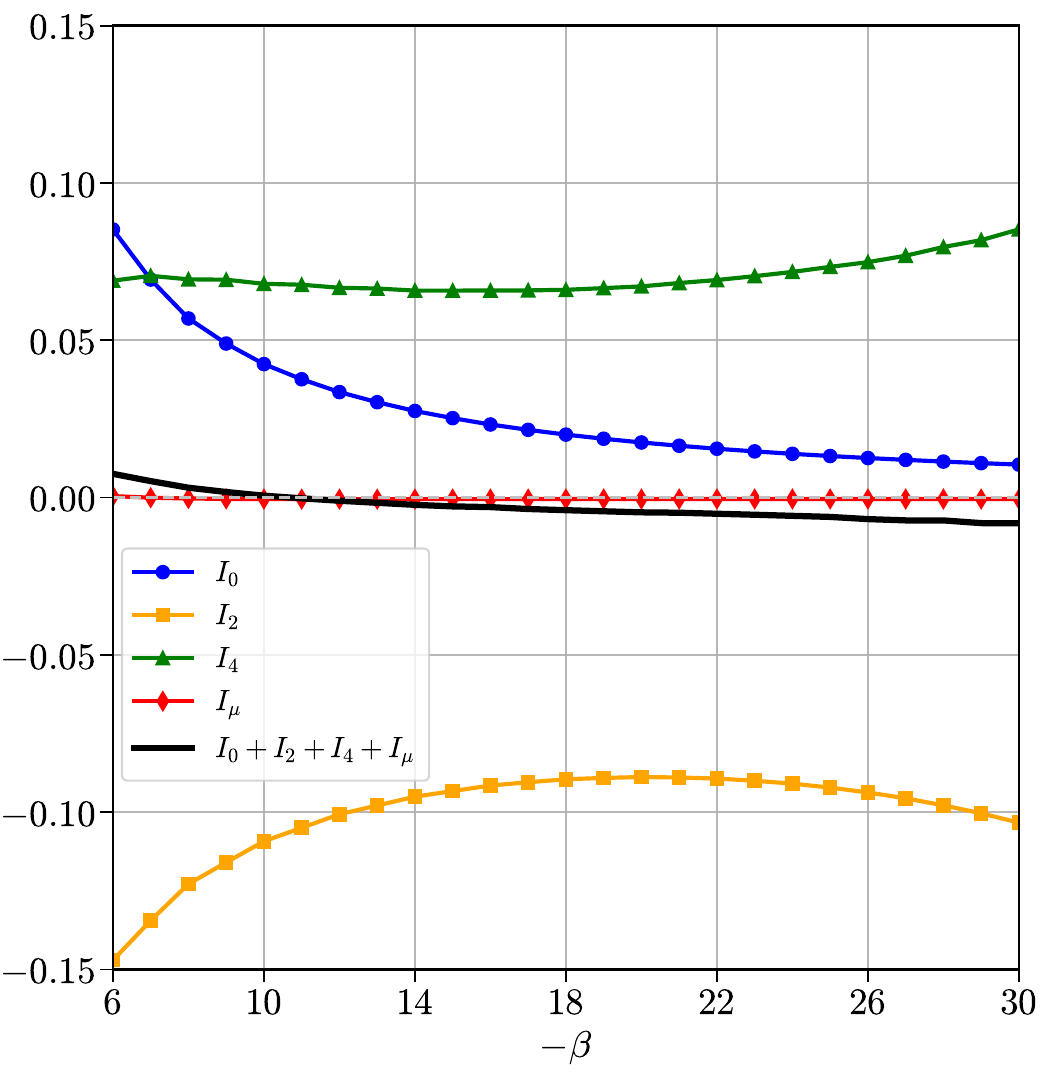}
    \caption{Dependence of the integral terms in Eq.~\eqref{eq: b integral} and their sum on $\beta$. It can be seen that the contribution of $I_\mu$ is negligible. Note that there is an overall factor of $|\beta|^2$ in Eq.~\eqref{eq: b integral}, hence $b_0$ grows with $\beta$ much faster than the sum here.}
    \label{fig: integral_terms}
\end{figure}
The contribution of the individual integrals $I_{i,\mu}$ in Eq.~\eqref{eq: b integral} can be seen in Figure~\ref{fig: integral_terms}. It is clear that the contribution of $I_\mu$ is ignorable. The other terms are comparable to each other, but the negative contribution of $I_2$ dominates as $\beta$ becomes more negative. This makes $b_0$ itself negative once we are below a certain $\beta$ value, which in turn leads to the appearance and dominance of first-order scalarization. 

There are several aspects of Figure~\ref{fig: integral_terms} that contributes to first-order scalarization being the norm at increasingly negative $\beta$. First, even though $I_0$ is a major positive contributor to $b_0$ at less negative $\beta$, its effect decreases quite steeply as $\beta$ becomes more negative, eventually rendering it ignorable. Secondly, $I_2$ and $I_4$ both grow in magnitude after a certain $\beta$ value, $-\beta \sim 20$ in Figure~\ref{fig: integral_terms}, essentially determining $b_0$. The negative term $I_2$ wins over to make $b_0$ negative, and consequently, scalarization becomes first-order. $I_2$ and $I_4$ have a different behavior at less negative $\beta$, they are not growing in magnitude. Our main task is revealing the reasons for first-order scalarization becoming the norm for $-\beta \gg 1$, so we will concentrate on the $-\beta \gtrsim 20$ region in this section, and discuss $-\beta \lesssim 20$ in Appendix~\ref{appx: decreasing I_2}.

\begin{figure*}
    \centering
    \includegraphics[width=\textwidth]{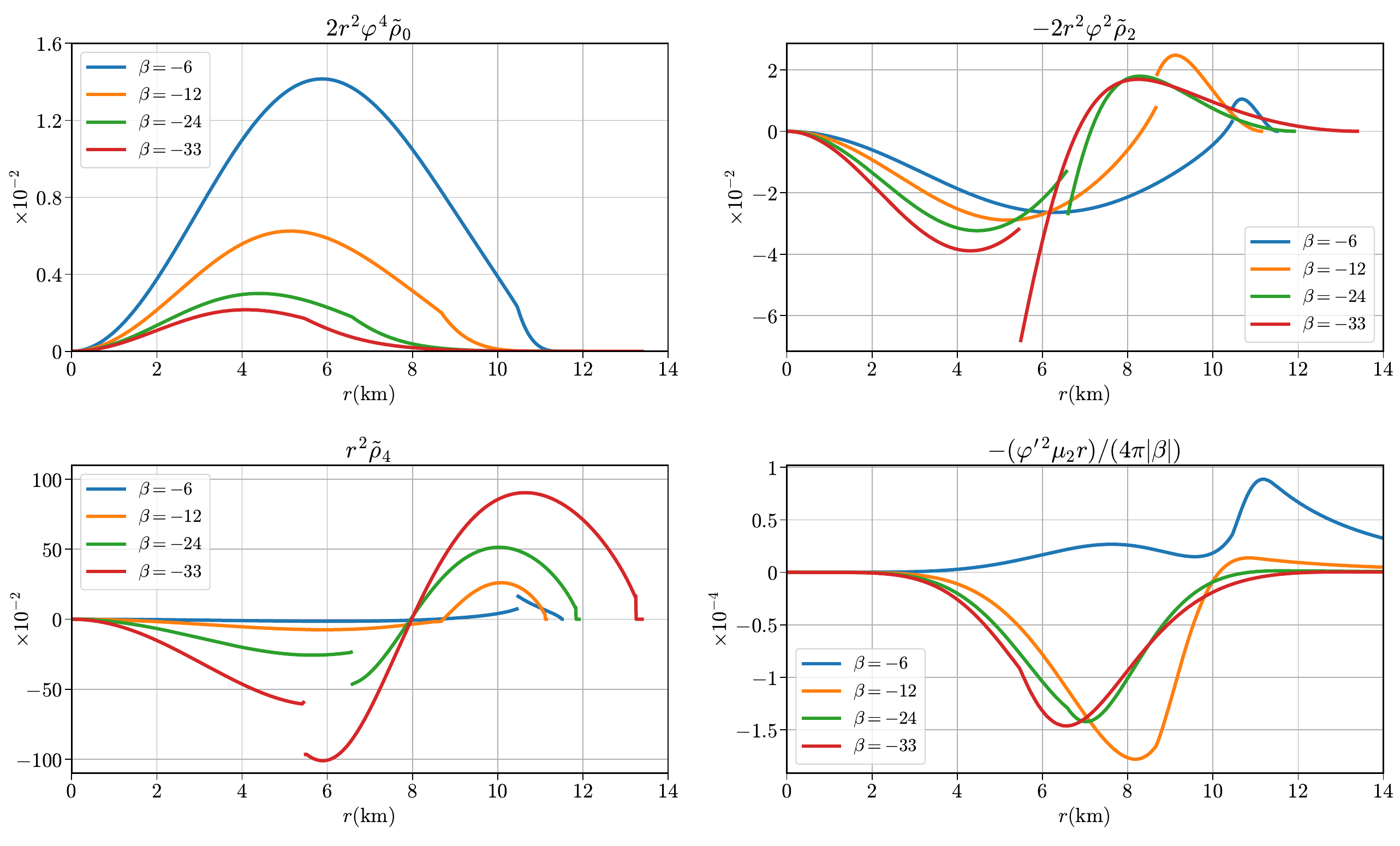}
    \caption{Each term in the $b_0$ integral of Eq.~\eqref{eq: b integral}. Note that there are discontinuities in the matter response terms $\rt_{2,4}$ which are expected due to the piecewise polytropic nature of our HB EOS~\cite{Read:2009yp}, see Appendix~\ref{appx: numerical}. The main feature is that as $|\beta|$ grows, the negative contribution from the $\rt_2$ term (top right) dominates over the positive contribution from the $\rt_0$ (top left) and the $\rt_4$ (bottom left) terms. Note that $r^2\rt_4$ reaches high magnitudes, but its positive and negative sections mostly cancel each other when integrated in Eq.~\eqref{eq: b integral}. The $\mu_2$ term (bottom right) has negligible effect compared to the others. Overall, the contribution of the $\rt_2$ term renders $b_0$ exceedingly more negative as $\beta$ attains more negative values, and first-order scalarization becomes the norm.}  
    \label{fig: b integral terms}
\end{figure*}

Overall, we observed the behavior in Figure~\ref{fig: integral_terms} for $-\beta \gtrsim 20$  to continue at least up to $\beta \sim -100$. $I_2$ and $I_4$ continue to grow in magnitude while the other two terms stay small, and $I_2$ continues to make $b_0$ increasingly negative for all sufficiently negative $\beta$, also helped by the overall $|\beta|^2$ factor in Eq.~\eqref{eq: b integral}. In the following, we will try to explain these observations by studying each of the integrands in $I_{i,\mu}$, which are plotted in Figure~\ref{fig: b integral terms}. Note that $\varphi$ depends relatively weakly on $\beta$, see Figure~\ref{fig: tachyon}. Hence, the $\beta$ dependence of $I_{0,2,4}$ essentially arises from $\rt_{0,2,4}$. Our main task will be understanding the behavior of these density terms.

Let us start with the most important factor in determining the sign of $b_0$: the sign of $I_2$ (we will discuss its magnitude later). For this, we need to study its integrand, $- 2 r^2 \varphi^2 \rt_2$ that is plotted on the top right panel of Figure~\ref{fig: b integral terms}. A star tends to become more compact when it scalarizes, which is an important factor in determining the sign of the matter response function. In rough terms, this can be understood as the star preferring to be scalarized, hence it would lose some of its energy by moving its matter into the deeper parts of the gravitational well. Since the total baryon mass is kept constant, higher compactness means the leading order change in energy density $\rt_2$ is positive near the core, but there is also a region near the surface where it is negative. The top right panel of Figure~\ref{fig: b integral terms} exactly follows this pattern. Note that the $r$ and $\varphi$ factors suppress $- 2 r^2 \varphi^2 \rt_2$ near the center and the surface of the star, hence the intermediate region dominates, making $I_2$ negative.

\begin{figure}
    \centering
    \includegraphics[width=\columnwidth]{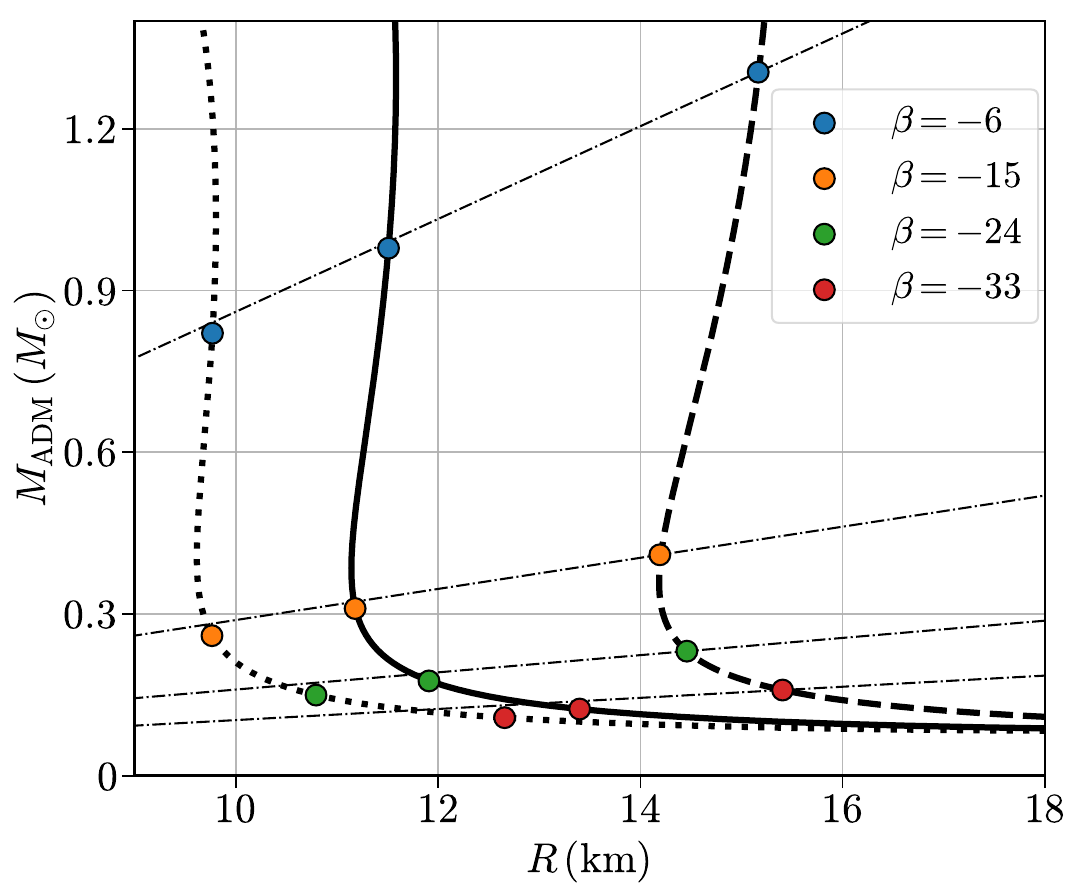}
    \caption{The mass-radius curves under GR for three increasingly stiffer EOS: 2B EOS (left, dotted line), HB EOS (middle, solid line), 2H EOS (right, dashed line)~\cite{Read:2009yp}. The linear, dash-dotted lines correspond to constant compactness values $\MADM/R$. The bifurcation points (marked with circles) move to lower stellar mass and compactness as $\beta$ becomes more negative. Except for Sec.~\ref{sec: eos modification}, we exclusively study the HB EOS, but the movement of the bifurcation point with $\beta$ is similar for all three.}
    \label{fig: mass radius}
\end{figure}

Next, let us explain why the positive term $I_0$ in Eq.~\eqref{eq: b integral} becomes less important as $\beta$ becomes more negative. Dependence of the integrand $2 r^2 \varphi^4 \rt_0$ on $\beta$ is in the top left panel of Figure~\ref{fig: b integral terms}. Observe that $\rt_0$, the unperturbed energy density of the star, becomes smaller in magnitude as $\beta$ becomes more negative, i.e. as $|\beta|$ increases. This is mainly because a more negative $\beta$ makes scalarization easier in the sense that the tachyon appears at lower stellar density values in Eq.~\eqref{eq: zero mode potential}. Consequently, the star at the bifurcation point becomes less massive and less compact, which can be seen in the stellar mass-radius plots of Figure~\ref{fig: mass radius}. This means $2 r^2 \varphi^4 \rt_0$ and its integral $I_0$ monotonically decreases as $\beta$ becomes more negative. Recall that this is a strictly positive term, so first-order scalarization is facilitated when its amplitude decreases.

Let us now turn back to $I_2$ and its integrand $- 2 r^2 \varphi^2 \rt_2$, and explain why the magnitude of their negative contribution to $b_0$ increases as $\beta$ becomes sufficiently negative. The underlying reason is again related to the movement of the bifurcation point to lower stellar masses in Figure~\ref{fig: mass radius}. However, $\rt_2$ is not directly affected by the decreasing stellar mass, unlike $\rt_0$. Rather, the increase in $\rt_2$ is related to the stiffness of the star going down. Realistic EOS are such that there is less stiffness at lower densities $\rt_0$, which is also the case for our choice. This means we can have a stronger matter response for the same perturbation, hence, $\rt_2$ naturally increases in amplitude as the bifurcation point moves to lower-mass stars with lower densities. Consequently, when $- 2 r^2 \varphi^2 \rt_2$ is integrated to form $I_2$, we expect it to grow in magnitude, which is the eventual behavior in Figure~\ref{fig: integral_terms}. This is the general trend, but there is also the transient behavior in the $-\beta \lesssim 20$ region of Figure~\ref{fig: b integral terms}, where $I_2$ decreases in magnitude with $-\beta$. We discuss this in Appendix~\ref{appx: decreasing I_2}.

To summarize the results so far, just using the top two panels of Figure~\ref{fig: b integral terms}  for now, the reason for $b_0$ eventually becoming increasingly negative as $\beta$ becomes more negative is (i) the decreasing positive contribution from $\rt_0$ and (ii) the increasing (in magnitude) negative contribution from $\rt_2$.

Figure~\ref{fig: b integral terms} has two other panels on the bottom row, which are the integrands of $I_4$ and $I_\mu$ in the $b_0$ integral~\eqref{eq: b integral}. Let us first discuss $\mu_2$ and the related integral $I_\mu$, which are relatively easier to interpret compared to $I_4$. As can be seen in the bottom right panel of Figure~\ref{fig: b integral terms}, $-\varphi'^2 \mu_2/(4\pi|\beta| r)$ becomes negative almost everywhere when $\beta$ becomes sufficiently negative, even before first-order scalarization appears. However, this term is ignorable compared to the ones in the other panels. This can be understood as a result of the suppression by the $1/|\beta|$ factor. $\mu_2$ can be expected to grow with $|\beta|$ due to the movement of the bifurcation point, similarly to $\rt_2$, but this growth is mild, hence canceled out by the $|\beta|$ in the denominator.

Lastly, we have the $I_4$ integral in Figure~\ref{fig: integral_terms}, which arises from integrating $r^2\rt_4$ in the bottom left panel of Figure~\ref{fig: b integral terms}. $\rt_4$ is clearly growing in magnitude with more negative $\beta$, similarly to $\rt_2$. The overall functional profile also shows that $\rt_4$ has the opposite sign to the leading matter response term $\rt_2$, being negative nearer to the core and positive near the surface. These can be understood when we recall that $\rt_4$ is the next-to-leading order response term to scalarization, at the $\phi_\text{c}^4$ order.

One main contributor to $\rt_4$ is the quartic term in the series expansion $A^4(\phi) = 1 + 2\beta \phi^2 + 2\beta^2 \phi^4+ \dots$ which controls the direct coupling between matter and the scalar field at the $\mathcal{O} (\phi_\text{c}^4)$ order. Note that $2\beta^2 \phi^4$ grows faster with $\beta$ and also has the opposite sign to the $\mathcal{O} (\phi_\text{c}^2)$ order term $2\beta \phi^2$ when $\beta$ is negative. This is likely the reason for the opposite sign of $\rt_4$ with respect to $\rt_2$, and also the fact that $\rt_4$ gets bigger in magnitude with $|\beta|$.

Interpreting $I_4$, that is, the contribution of $r^2 \rt_4$ to Eq.~\eqref{eq: b integral}, is slightly harder than the previous terms, and our understanding is not as clear as the previous cases. This is because $r^2 \rt_4$ has large positive and negative sections, hence the sign of its integral $I_4$ is not obvious. We consistently observed that this term makes a positive contribution to $b_0$ as is evident in Figure~\ref{fig: integral_terms}, hence it tries to inhibit first-order scalarization. Moreover, $I_4$ eventually becomes the main positive contribution to $b_0$ when $\beta$ becomes sufficiently negative, yet it is still dominated by the negative contribution from $I_2$. One reason we have identified for this observation is the lack of the $\varphi$ factor in $r^2 \rt_4$ (the integrand of $I_4$), which is present in $- 2 r^2 \varphi^2 \rt_2$ (the integrand of $I_2$). Without the slight suppression arising from $\varphi \sim 0.5$ near the stellar surface, the negative and positive portions of $r^2 \rt_4$ become of similar magnitude, see Figure~\ref{fig: b integral terms}, and they cancel each other more efficiently when integrated, unlike the case for $- 2 r^2 \varphi^2 \rt_2$.

Let us summarize our understanding. Existence of first-order scalarization requires that the spacetime and matter in it respond to the growth of the scalar field. Dominance of first-order scalarization at more negative $\beta$ indicates that matter response terms that have negative contributions to $b_0$ in Eq.~\eqref{eq: b integral} should dominate over the others as $\beta$ becomes more negative. This exact scenario is realized for the leading matter response term $\rt_2$ and its related integral $I_2$, which contribute negatively to $b_0$. More negative $\beta$ moves the star at the bifurcation point to lower stellar masses, hence decreasing its average stiffness. This makes the star react more readily to the scalar field, increasing $\rt_2$ while the unperturbed density $\rt_0$ decreases. The behavior of the next-to-leading response term $\rt_4$ and its related integral $I_4$ are more complex, and it is especially not easy to understand why $I_4$ and $I_2$ are comparable in magnitude. Yet, there are reasons to expect $I_4$ to be subdominant to $I_2$ such as the $\varphi$ terms in the integrand. Overall, as $\beta$ becomes more negative, the interplay of all these contributions make $b_0$ more negative as well. This is the hallmark of first-order scalarization, whose prominence in this $\beta$ regime we set out to understand.

\section{The predictive power of the energy function}
\label{sec: predictions}
We stated our main aim in developing the energy function framework to be gaining insight into how the phase transition changes for different models of scalarization, which can potentially circumvent costly numerical computations. In this section, we demonstrate this by providing predictions about what happens to scalarization when we deviate from the original DEF model of scalarization. As in the previous cases, our main focus will be explaining how the order of the transition behaves.

\subsection{Effect of the conformal coupling function $A(\phi)$}
\label{sec: coupling modification}
So far we have exclusively considered the original conformal coupling function $A(\phi)$ of the DEF model~\cite{Damour:1993hw} which rendered our theory a single-parameter one. One central aspect of this choice has been that many of our results are direct consequences of the series expansion of this function, or rather, its fourth power
\begin{equation}\label{eq: Aphi expansion DEF}
    A^4(\phi) = e^{2\beta\phi^2} = 1 + 2\beta \phi^2 + 2\beta^2 \phi^4 + \cdots.
\end{equation}
For example, one of the most consequential terms in the $b_0$-integral formula~\eqref{eq: b integral}, $2 r^2 \varphi^4 \rt_0$ can easily be traced to arise from matter coupling to the quartic coefficient in the expansion of $A^4(\phi)$, i.e., $2\beta^2 \phi^4$.

However, there is nothing essential about the above functional form, and in principle $A(\phi)$ can be changed to depend on various other parameters. The essence of scalarization is a tachyonic instability, whose existence most directly depends on the quadratic term in the expansion of $A$,\footnote{There is also the concept of \emph{nonlinear scalarization} which can occur in the absence of a quadratic term~\cite{Doneva:2021tvn}, but we will stay focused on the more common, tachyon-based form here.} so we can modify higher order terms to obtain other models of scalarization. For example, we can modify the fourth order term via a new parameter:
\begin{equation}\label{eq: Aphi expansion}
    A^4(\phi) = e^{4 \left(\frac{\beta}{2}\phi^2- \frac{\gamma}{24} \phi^4 \right)} = 1 + 2\beta \phi^2 + \left( 2\beta^2 -\frac{\gamma}{6}\right) \phi^4 + \cdots ,
\end{equation}
where we kept the overall exponential form for ease of comparison to our previous results. Such models can have radically different phase transition properties depending on the value of $\gamma$. For example, the $\rt_0$ term in the $b_0$-integral~\eqref{eq: b integral} would change as
\begin{equation}\label{eq: rt0 delta}
    2\varphi^4 \rt_0 \to  \left( 2 -\frac{\gamma}{6|\beta|^2}\right) \varphi^4 \rt_0.
\end{equation}
Recall that this term was the dominant positive contributor to $b_0$ at lower $|\beta|$, and made it possible that scalarization is second-order. With the introduction of $\gamma$, we can, for example, flip the sign of this term to make sure that $b_0$ is always negative, and second-order phase transition is not observed for any value of $\beta$.

\begin{figure}
    \centering 
    \includegraphics[width=\columnwidth]{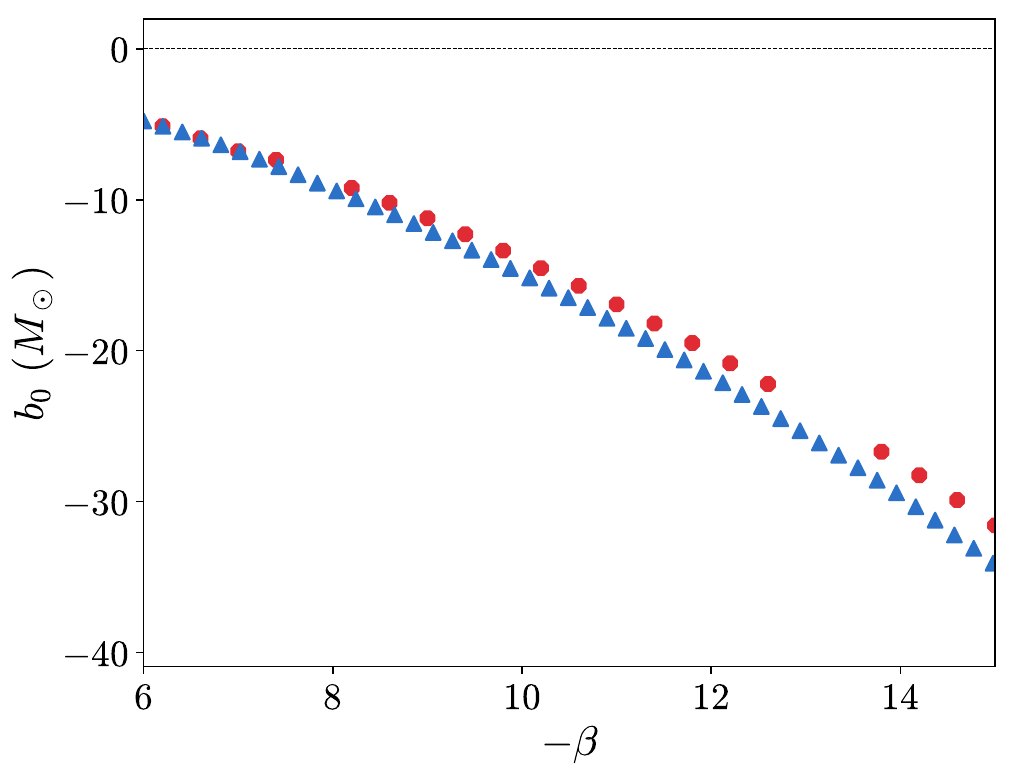}
    \caption{$b_0(\beta)$ for scalarization in a model with $A(\phi) = \exp({\frac{1}{2}\beta\phi^2 - \frac{1}{2}\beta^2\phi^4})$. This plot is an analog of Figure~\ref{fig: b0 vs beta}, that is, the red circles are obtained by numerical computations, and the blue triangles from the energy function framework. A qualitative study of the energy function already predicted this $A(\phi)$ to exclusively feature first-order scalarization (negative $b_0$) even before the computation of the blue curve (see the main text), which is confirmed by the red data points.}
    \label{fig: b_vs_beta_coupling_2}
\end{figure}

We can concretely see this in Figure~\ref{fig: b_vs_beta_coupling_2}, which is the analog of Figure~\ref{fig: b0 vs beta}, but for the new conformal scaling function $A(\phi) = \exp({\frac{1}{2}\beta\phi^2 - \frac{1}{2}\beta^2\phi^4})$. The specific value of $\gamma = 12\beta^2$ is chosen such that the $\rt_0$ term, Eq.~\eqref{eq: rt0 delta}, vanishes (it is not made negative). As  a consequence, $b_0(\beta)$ is negative for all cases of scalarization, and the phase transition is first-order for any value of $\beta$.

Figure~\ref{fig: b_vs_beta_coupling_2} is just one example of what can happen for a more generic $A(\phi)$, the energy function framework provides a strong predictive power in general. All else being equal, the effect of $A(\phi)$ on the order of the phase transition is mainly determined by the quartic term in $A^4(\phi)$. If we make this term more positive,\footnote{Note that we also need to ensure that our energy function provides global stability by being bounded from below, so we should be careful when considering different functional forms of $A(\phi)$.} then we can also increase the range of $\beta$ for which $b_0>0$ and scalarization is second-order, however there is still a  move towards first-order scalarization with increasingly negative $\beta$. Overall, this provides a clear example of what we set out to do: having a predictive framework that can inform us about the phase transition properties of scalarization without the need to directly compute a large number of neutron star configurations.

Lastly, we also want to comment on the importance of $A(\phi)$ for astrophysical relevance. Note that so far, the bifurcation point and the related metastability phenomena in first-order scalarization occurred at stellar masses considerably lower than one solar mass, which is well below most current measurements of neutron star masses~\cite{Freire:2024adf}. However, for an $A(\phi)$ that further facilitates first-order scalarization, like the one in Figure~\ref{fig: b_vs_beta_coupling_2}, the bifurcation points would be the same as the case in the DEF model~\eqref{eq: Aphi} since we have not changed the quadratic term. Hence, for such modified $A(\phi)$, we can encounter metastability and possible  transitions between locally stable configurations at much higher masses, say $\sim 1.5 M_\odot$, which is where the bifurcation point is for less negative $\beta$ such as $\beta\ \sim -5$. This is an important point for relating our theoretical findings to observational signals in the future.

\subsection{Effect of the scalar field mass and self interaction}
\label{sec: mphi modification}
%
\begin{figure} 
    \centering 
    \includegraphics[width=\columnwidth]{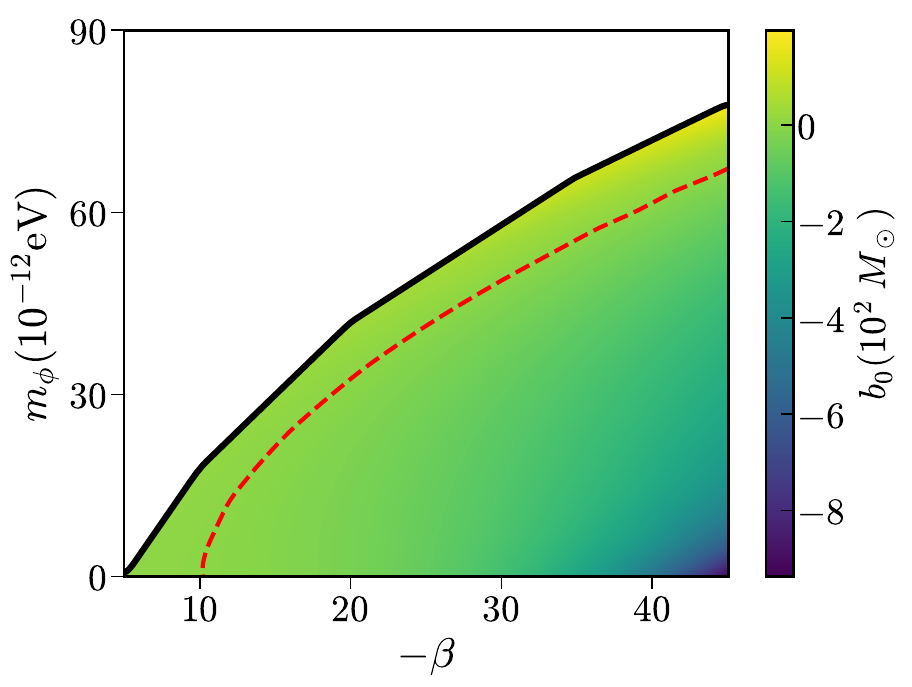}
    \caption{$b_0$ values computed via the energy function framework, i.e. Eq.~\eqref{eq: b integral}, when a scalar mass term $\mphi$ is introduced in addition to the scalar coupling parameter $\beta$. As predicted by the energy function, increasing $\mphi$ moves $b_0$ towards positive values, which means scalarization becomes second-order. The dashed red line is the $b_0=0$ contour. This plot is an analog of Figures~\ref{fig: b0 vs beta} and~\ref{fig: b_vs_beta_coupling_2}, but on a two-dimensional $(\beta, \mphi)$ parameter space. Like the previous figures, the numerical values closely agree with what numerical computations via Eq.~\eqref{eq: fit b_0} provide (comparison not shown).}
    \label{fig: b of massive stt}
\end{figure}
We have considered massless scalars so far, which was the original case in the DEF model, but this model is now mostly ruled out by astrophysical observations~\cite{Zhao:2022vig}. However, the essential aspects of scalarization are not changed radically for a massive scalar, which also are immune to these observational tests. Moreover, this is still the case for relatively small scalar masses which virtually leave the scalarized neutron star configurations the same as the ones for massless scalars~\cite{Ramazanoglu:2016kul}. Overall, it is interesting to understand what happens when we add a mass term $\mphi$ to the scalar field both in terms of astrophysical relevance and also as a natural theoretical extension.

We should add for the sake of transparency that we already know \emph{how} a scalar mass term affects the phase transition, which was reported in \textcite{Unluturk:2025zie}, however, there has not been an explanation for the observed behavior so far. Hence, what we aim for can more aptly be called a ``postdiction'' rather than a prediction. We will not assume anything about the numerical findings in the subsequent discussion, and see that our framework can be adapted to massive scalars as well.

Mathematically, the mass term is added by the change 
\begin{align}
\label{eq: mphi}
    - 2g^{\mu\nu} \nabla_\mu\phi\nabla_\nu\phi \to - 2g^{\mu\nu} \nabla_\mu\phi\nabla_\nu\phi - 2\mphi^2\phi^2
\end{align}
in the original action~\eqref{eq: stt action}. It shows up as a simple constant coefficient in the scalar field equation of motion
\begin{equation}
\label{eq: scalar field eqn mphi}
    \square_g\phi = \left(-8\pi A^4 \frac{\d\left(\ln A\right)}{\d\left(\phi^2\right)} \tilde{T} + \mphi^2 \right)\phi \equiv m_\text{eff}^2 \phi,
\end{equation}
making the effective mass squared strictly more positive, hence, less tachyonic, cf. Eq.~\eqref{eq: scalar field eqn}. Thus, increasing scalar mass ``weakens'' scalarization in broad terms, and extremely high values make it altogether impossible~\cite{Ramazanoglu:2016kul}. 

How does $\mphi$ affect the phase transition? The first point to note is that there is no change in the formula for $b_0$, Eq.~\eqref{eq: b integral}. This suggests that the increasing dominance of first-order scalarization as $\beta$ becomes more negative is also expected for a constant $\mphi$ line on the $(\beta, \mphi)$ plane. This is indeed the case as we can observe in Figure~\ref{fig: b of massive stt}.

What about changing $\mphi$ while keeping $\beta$ constant? Even though the $b_0$ formula~\eqref{eq: b integral} is the same, the terms in the formula such as $\rt_{0,2,4}$ vary with $\mphi$ because the bifurcation point moves. We saw that $\mphi$ makes it harder for a star to scalarize, which means all else kept the same, the first star to scalarize, the one at the bifurcation point, needs to have a higher density profile and higher total stellar mass, cf. Eq.~\eqref{eq: zero mode potential}. In other words, the bifurcation points in Figure~\ref{fig: mass radius} move up with increasing $\mphi$.

We already discussed the effects of moving the bifurcation point in Sec.~\ref{sec: explaining first-order}. In that case more negative $\beta$ moved us towards less massive stars, i.e., lowered the bifurcation point on Figure~\ref{fig: mass radius}. Hence, we can quickly see that increasing $\mphi$ is expected to have the opposite effect: it moves $b_0$ towards more positive values, and facilitates second-order scalarization. This is also confirmed in Figure~\ref{fig: b of massive stt}.

To summarize, more negative $\beta$ at a fixed scalar mass moves scalarization towards being first-order for any $\mphi$. Increasing $\mphi$ while keeping $\beta$ constant behaves in the opposite way, moving scalarization away from being first-order. Though, to be strict, we should add that increasing $\mphi$ and decreasing $|\beta|$ are not equivalent in all aspects due to their different effects on the matter distribution. The combined result of $\beta$ and $\mphi$ on $b_0$ can be seen in Figure~\ref{fig: b of massive stt}, which confirms our qualitative explanation.

We can extend our reasoning here to more general scalar field potentials $V_\phi$ that generalize a mass term:
\begin{align}
    - 2g^{\mu\nu} \nabla_\mu\phi\nabla_\nu\phi \to - 2g^{\mu\nu} \nabla_\mu\phi\nabla_\nu\phi - V_\phi(\phi).
\end{align}
For example, we can add one more term while keeping the $\phi \to -\phi$ symmetry~\cite{Staykov:2018hhc}
\begin{align}
    V_\phi(\phi) = 2\mphi^2 \phi^2 +\lambda \phi^4,
\end{align}
where $\lambda>0$ to ensure that the energy is bounded from below. $\lambda$ does not affect the bifurcation point since it does not contribute to the linear differential equation~\eqref{eq:schrodinger} for the zero mode. However, it is straightforward to see that the formula for $b$ has a novel positive contribution proportional to $\lambda$, hence increasing this term pushes scalarization towards being second-order.

\subsection{Effect of the EOS}
\label{sec: eos modification}
Even though it is not a theory parameter, the EOS for nuclear matter is another unknown, and it affects scalarization. The main relevant property of the EOS for us is its stiffness, which directly manifests as a higher speed of sound as well. A stiff EOS also typically provides higher pressure at a given energy density compared to a soft one. We will see that despite the unrelated physical origin, the effects of EOS stem from similar mechanisms to the theory parameters $\beta$ and $\mphi$. Namely, EOS also changes the bifurcation point. However, unlike the cases considered so far, EOS also changes the whole GR solution curve on the stellar mass-radius plane as can be seen in Figure~\ref{fig: mass radius}. Hence, the overall effect of EOS on the scalarization phase transition depends on the interplay of these two factors.

First, let us understand what happens when the EOS becomes more stiff. Effective mass of the scalar in Eq.~\eqref{eq: scalar field eqn} is dependent on the trace of the stress-energy tensor $\tT$, which in turn changes with EOS. We ignored the pressure terms as an approximation in Sec.~\ref{sec: scalarization as a phase transition}, and disregarded it completely for the toy star, but the exact expression is $\tT = -\rt + 3\pt$, and $\pt$ becomes an important factor especially near the core of more massive stars. When we linearize the field equations to search for a tachyon, the potential well~\eqref{eq: zero mode potential} is therefore shallower, so the zero mode appears at more massive and typically more compact stars. In other words, the bifurcation points shift upwards with increased stiffness, which is clearly observed in Figure~\ref{fig: mass radius}. This is very similar to the effect of increasing $\mphi$, hence the result is also similar. So, $b_0$ moves to more positive values and second-order scalarization is facilitated due to this effect. However, this is not the only effect of the EOS.

The second major result of stiffness is modifying the whole neutron star solution curve, as can be clearly seen in Figure~\ref{fig: mass radius}. Stiffer EOS shift the whole mass-radius curve to higher radius values. This has a lowering effect on the compactness of the stars at the same ADM masses, which leads to more negative $b_0$ and first-order scalarization. 

Overall, as the stiffness of the EOS increases, the two main effects above compete with each other. Unlike our previous analysis for $\beta$, $\mphi$ and the form of $A(\phi)$, it is not possible to have a sharp answer without knowing the details of the EOS. However, Figure~\ref{fig: mass radius} also reveals the interesting fact that even though the stellar masses and the radii of the bifurcation points change, their compactness is almost constant. That is, for a given $\beta$, the bifurcation points all lie very close to the same compactness line. This indicates that the two opposing effects of EOS stiffness above approximately cancel each other. This also curiously follows the behavior of a toy star, where the onset of scalarization depended only on compactness, and not on the mass or radius, see Eq.~\eqref{eq: compactness} and the subsequent discussion. This behavior is reminiscent of universal relations for neutron stars~\cite{Yagi:2016bkt}.

\begin{figure}
    \centering
    \includegraphics[width=\columnwidth]{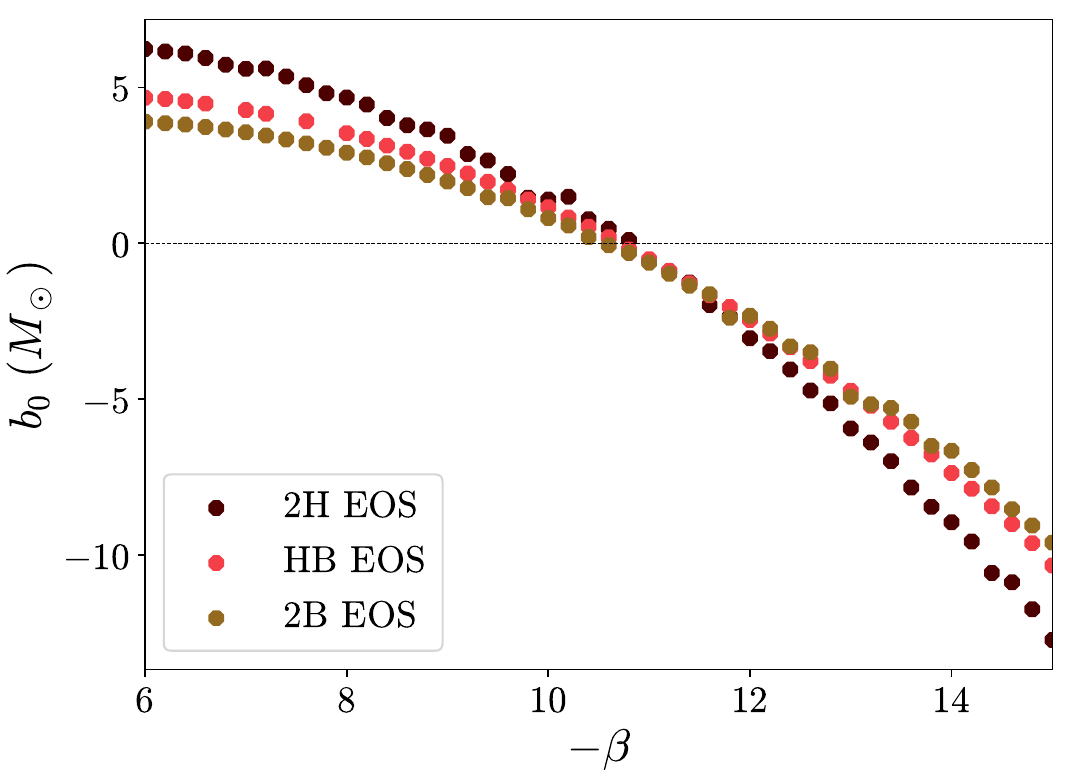}
    \caption{$b_0(\beta)$ calculated numerically via Eq.~\eqref{eq: fit b_0} for the increasingly stiff 2B, HB and 2H EOS~\cite{Read:2009yp}. The difference between EOS is small and it also changes sign with $\beta$. Increased stiffness favors more positive $b_0$ and second-order scalarization at lower $|\beta|$, and more negative $b_0$ and first-order scalarization at higher $|\beta|$. This is likely due to the interplay of how the bifurcation point moves on a given mass-radius curve versus how the whole mass-radius curve itself is shifted to higher radii for increasingly stiff EOS in Figure~\ref{fig: mass radius}.}
    \label{fig: b0 EOS}
\end{figure}
A quantitative comparison can be seen in Figure~\ref{fig: b0 EOS}, where we plot the numerical computations for $b_0(\beta)$ for three EOS from the same piecewise polytropic family as our HB EOS: the softer 2B EOS\footnote{The maximum neutron star mass mass for 2B EOS is below many measurements, hence this EOS is mostly ruled out. However, it is still useful to test our understanding of the effect of EOS stiffness on scalarization.}, and the stiffer 2H EOS~\cite{Read:2009yp}. The differences between the EOS are muted, which agrees with our discussion above. However, the differences due to EOS stiffness are still observable, and curiously $\beta$-dependent. Stiffer EOS are more positive at lower $|\beta|$ as if the effect from the movement of the bifurcation point on a given solution curve dominates, but this pattern is reversed at higher $|\beta|$ as if the whole shift of the mass-radius curve to higher radii is the dominant effect. We do not have an explanation for this particular trend.

The main lesson is that we can understand the magnitude of the effect of EOS on the order of scalarization: it is small. However, there is still structure in this effect in terms of its dependence on $\beta$, which is harder to surmise. We should also add that we restricted our investigation to an EOS family that behaves in the typical manner expected from nuclear matter. Considering nuclear phase transition and more exotic behavior might change this picture.

\section{Conclusions}
\label{sec:conclusions}
The aim of this manuscript was deriving the energy function of a neutron star under spontaneous scalarization starting from the first principles, that is, the exact field equations of the theory. As we summarized, all the scalarized and unscalarized solutions of a given scalar-tensor theory, i.e.~the phase diagram, can be numerically computed given enough resources. Before the current study, we also knew how to explain this structure in a phenomenological approach by positing an energy ansatz that explains all of the qualitative, and some of the quantitative, features of the phases. However, such an approach does not provide a fundamental understanding of why we see a first-order transition in some cases, and a second-order one in others. It does not even tell us what would happen in a slightly different theory, for example, when the coupling function is modified, either.

To address the above problem, we have derived the energy of a neutron star as a function of the scalar field strength using perturbation theory. This provides the dependence of the solution space and the phase transition picture on theory parameters as well as the EOS, which elucidates the underlying mechanisms of the scalarization process. Our framework explains recent numerical observations like the dominance of the first-order phase transition on the parameter space of the DEF theory. Beyond providing a bottom-up reasoning for the existing computational results, we also demonstrated the predictive power of our first-principles approach by considering the phase transition implications of various terms in the scalar tensor theory, such as the conformal coupling function $A(\phi)$ and scalar mass $\mphi$.

The response of matter to the scalar field turned out to be a major factor in determining the phase transition properties of scalarization. If one uses a fixed background metric and matter profile, scalarization is always second-order in the original DEF model. In general, the order only depends on the quartic term in the series expansion of $A(\phi)$. However, we demonstrated that the modifications to the matter profile due to the growing scalar field provides contributions that favor first-order scalarization. Moreover, these contributions grow with increasingly negative $\beta$, explaining why this part of the parameter space exclusively features first-order scalarization.  Specifically, we identified the change in the mass and stiffness of the star where the tachyon first appears, the one at the bifurcation point in our terminology, to be the culprit of the phase transition trends as the coupling parameter $\beta$ changes.

One would naturally expect the matter response to play some role in scalarization, but our findings are far from obvious. For example, the bifurcation point can be found solely by using a linearized scalar field equation on a fixed background. Such an analysis can directly determine whether scalarization exists or not, and has been popular in the literature, rightly so in our opinion. Thus, we can learn some aspects of scalarization while ignoring the nonlinear terms like the matter response. However, understanding scalarization as a phase transition requires information about the final scalarized solutions, not just the existence of the tachyonic instability. This requires a deeper analysis of the nonlinear effects, which is likely one of the reasons for the relatively late discovery of the prevalence of first-order scalarization.

As we have mentioned, the order of scalarization has important consequences for potential astrophysical observations. Existence of metastable configurations and transitions between them and the ground state configurations can lead to new signals, some of which have already been investigated~\cite{Kuan:2022oxs}. Moreover, any study that tries to compare neutron star data to scalarization has to be careful about determining which of the locally stable states (if there are more than one) is assumed to be realized in our universe~\cite{Tuna:2022qqr,Demirboga:2023ktt}.\footnote{The first systematic study of metastable scalarized solutions were performed in such a data analysis effort, namely \textcite{Tuna:2022qqr}, which was the inspiration for the later phase transition analysis of \textcite{Unluturk:2025zie}.} Hence, in the case of first-order scalarization, data analysis requires additional attention.

We also want to briefly discuss the range of stellar mass values where the metastable configurations are possible, i.e. the region around the bifurcation point for the red curve of Figure~\ref{fig: Mb vs rho_c}. This is the region where first-order scalarization is truly different from second-order, hence neutron stars with masses in this interval are the prime targets for any new observation channels. However, in the case of Figure~\ref{fig: Mb vs rho_c} and many others, the interesting metastable solutions only occur below one solar mass, well below most current neutron star mass measurements~\cite{Freire:2024adf}.

We believe that the unique aspects of first-order scalarization are still astrophysically relevant for a variety of reasons. Firstly, ever decreasing masses of neutron stars are being observed, a recent $0.77M_\odot$ example being the least massive one so far~\cite{Doroshenko2022LightNS}. This is already similar to the mass of metastable neutron stars for certain favorable choices of $\beta$ and $\mphi$~\cite{Unluturk:2025zie}. Secondly, there are also proposals that allow even less massive neutron stars like formation channels in collapsar disks and partial disruption events~\cite{Metzger:2024ujc,Stephens:2011as,East:2011xa}. Thirdly, and most importantly in our opinion, a main lesson from the current study is that scalarization can behave radically differently as a phase transition when we deviate from the DEF model. Hence, we expect distinct models of scalarization to allow metastable configurations (and transitions from and to them) at much higher stellar masses. The modification of the quartic term in $A(\phi)$ in Sec.~\ref{sec: coupling modification} is one simple example. 

Considering different $A(\phi)$ is related to the physical meaning of our scalar-tensor theory action, which can be best understood as an effective field theory for modifications to GR due to some yet-unknown scalar field. The exact functional form of $A(\phi)$ is not physically important on its own in many cases, it is a placeholder used by theoreticians. However, the implied scalar coupling structure arising from the Taylor expansion of $A(\phi)$ is meaningful as we have seen. Any potential observable related to mere existence of scalarization would inform us about the quadratic coupling, since this is what determines whether there is a tachyonic instability in the first place~\cite{Doneva:2022ewd}. Similarly, the phase transition order of scalarization is directly dependent on the quartic coupling to the scalar field, hence observables related to this feature would probe higher orders in the effective field theory expansion.

The case we concentrated on, neutron stars in the DEF model of scalarization, is still restrictive if we recall that scalarization is quite a generic phenomenon within scalar-tensor theories~\cite{Doneva:2022ewd}. However, just as the simplified case of the toy star provided the template for understanding realistic stars, we expect our current tools to be adaptable to other models of scalarization. A prime target is theories where the scalar field couples to the Gauss-Bonnet invariant and black holes can also scalarize in addition to neutron stars~\cite{Doneva:2017bvd,Silva:2017uqg}. Second-order transitions have dominated the investigation of such models similarly to the case of the DEF model, but initial studies show that first-order scalarization is possible depending on the coupling parameters~\cite{Herdeiro:2026sur}. Unlike the DEF model, first-order scalarization does not dominate the theory parameter space in this case, and there are fundamental differences between black holes and neutron stars. Nevertheless, the scalarization processes have more similarities than differences, hence we expect our methods to be generalized to black hole scalarization in the future.

\acknowledgments
We thank O\u{g}uzhan K. Yamak for his various comments on the manuscript and the numerical computations. F.M.R was supported by the Scientific and Technological Research Council of Turkey (T\"UB\.ITAK) Grant Number 122F097 during part of this study.

\appendix
\renewcommand{\theequation}{A.\arabic{equation}}
\setcounter{equation}{0}
\numberwithin{equation}{section}

\section{Perturbative solution of the TOV equations}
\label{appx: response contribution}

Here, we explicitly provide the systems of differential equations that appear at each perturbative order when we insert the series expansion~\eqref{eq: tov perturbative expansion} into the exact nonlinear TOV system~\eqref{eq: tov full}.

The zeroth order, $\mathcal{O}(\phi_\text{c}^0)$, is the case the scalar field is completely ignored, which is the GR solution at the bifurcation point in Figure~\ref{fig: Mb vs rho_c}:
\begin{subequations}
\label{eq: gr tov}
\begin{align}
    \mu_0'(r) &= 4\pi r^2 \rt_0, \\
    \nu_0'(r) &= \frac{2(\mu_0 + 4\pi r^3 \pt_0)}{r(r - 2\mu_0)},\\
    \pt_0'(r)   &= -\frac{\rt_0 + \pt_0}{2} \nu_0'.
\end{align}
\end{subequations}
These equations can be integrated out for a given central pressure or density. Recall that it is not possible to find direct analytical solutions to such nonlinear systems, so we need to perform numerical computation, which means our results will be \emph{semi-analytical} unlike the case of the toy star. Yet, we can surmise the properties of the phase transition by understanding the basic physics of the functions being integrated, i.e. without the numerical details, see Sec.~\ref{sec: explaining first-order}.

In the next order, we have the $\mathcal{O}(\phi_\text{c}^2)$ system\footnote{Note that there is no equation for $\varphi$ here. Recall that we assume $\varphi$ to be determined after Eq.~\eqref{eq:schrodinger}, and calculate the response of the other functions to it.}
\begin{subequations}
\label{eq: tov2}
\begin{align}
    \mu_2'(r) &= 4\pi r^2 \rt_2(r) - 8\pi r^2\rt_0\varphi^2+\frac{1}{2|\beta|}r(r-2\mu_0)\varphi'^2, \\
    \nu_2'(r) &= \frac{1}{|\beta|}r\varphi'^2 - \frac{16\pi r^2 \varphi^2\pt_0}{r-2\mu_0} + \frac{8\pi r^2 \pt_2}{r-2\mu_0} \nonumber\\
    & \qquad + \frac{2 \mu_2/r + 2\nu_0'\mu_2}{r-2\mu_0}, \\
    \pt_2'(r) &= -(\rt_0 + \pt_0)(\nu_2'/2 - \varphi\varphi') - (\rt_2 + \pt_2)\nu_0'/2 .
\end{align}
\end{subequations}

And finally at the quartic $\mathcal{O}(\phi_\text{c}^4)$ order we have\\
\begin{subequations}
\label{eq: tov4}
\begin{align}
    \mu_4' &= 4\pi r^2 \rt_4 - 8\pi r^2 \varphi^2 \rt_2 + 8\pi r^2 \varphi^4 \rt_0 -\frac{1}{|\beta|} r \mu_2 \varphi'^2, \\
    \nu_4' &= \frac{1}{r-2\mu_0}\bigg(
    8\pi r^2 \tilde{p}_4 
    - 16\pi r^2 \varphi^2 \tilde{p}_2 + 16\pi r^2 \varphi^4 \tilde{p}_0 \notag\\
    & \quad\quad \quad \quad \quad \quad   
    + \frac{2\mu_4}{r} + 2\nu_0' \mu_4 + 2\nu_2' \mu_2
    - \frac{2}{|\beta|}r \mu_2 \varphi'^2
    \bigg),\\
    \pt_4' &=
    -(\tilde{\rho}_0 + \tilde{p}_0)\frac{\nu_4'}{2} - (\tilde{\rho}_2 + \tilde{p}_2)\left(\frac{\nu_2'}{2} 
    - \varphi \varphi'\right) \nonumber\\
    & \quad\quad \;- (\tilde{\rho}_4 + \tilde{p}_4)\frac{\nu_0'}{2}.
\end{align}
\end{subequations}

Note that the relationship between $\pt_2$ and $\rt_2$ is given by the derivatives of the EOS relationship $\rt(\pt)$, similarly for $\pt_4$ and $\rt_4$. This and various other subtleties to solving these equations numerically can be found in Appendix~\ref{appx: numerical}.

\section{Numerical methods}
\label{appx: numerical}
All numerical equations in Sec.~\ref{sec: Energy Function Framework For a Realistic Neutron Star} were solved using explicit finite difference integrators, mainly a custom implementation of the fourth-order Runge-Kutta (RK) method and the RK4(5) implementation in the \texttt{numpy} module of Python~\cite{harris2020array}. We followed the numerical methods of \textcite{Unluturk:2025zie} in the cases where we needed to compute an exact stellar structure by integrating the fully nonlinear system of TOV equations in Eq.~\eqref{eq: tov full}, for example to validate our new energy function framework.

The zeroth-order equations~(\ref{eq: gr tov}) which correspond to a GR solution are integrated from $r=0$ out as is usual for the TOV equations, where we slightly perturb the central energy density at the bifurcation point to ensure the presence of a tachyonic instability at the next order. The tachyonic instability is then identified by computing the smallest eigenvalue of the $\mathcal{O}(\Delta h^2)$ discretized operator given in Eq.~(\ref{eq:schrodinger}). The boundary conditions preserve the symmetry of the discretized operator, and hence its Hermiticity. In the perturbative regime, there is a single eigenvalue slightly below zero, whose corresponding eigenvector determines the profile of the tachyonic mode $\varphi(r)$. We verified the convergence of these finite difference computations.

The equations in the higher perturbative orders for the metric corrections, Eqs.~\eqref{eq: tov2} and~\eqref{eq: tov4},  are solved using a shooting method. We again integrate the system from the center out using a range of trial values $\pt_2(0)$ at the second perturbative order. We determine the correct initial values at the origin by enforcing the baryon mass conservation. Let us further detail this subtle point. 

Recall that we are interested in an expansion of the total energy of the spacetime in terms of the strength of scalarization $\phi_\text{c}$, that is, Eq.~\eqref{eq: ADM mass expansion}. In this expansion, we fix the baryon mass $\Mb$, so we are comparing different possibilities for scalarization on a star with a constant number of baryons. However, when the matter profile responds to scalarization and becomes perturbed via the terms $\pt_{2,4}$, these perturbations in general can change the baryon number from what we have at the zeroth order $\pt_0$, even though they are not supposed to. Correspondingly, we also need an initial value at the center of the star to integrate the $\pt_{2,4}$ equations in Eqs.~\eqref{eq: tov2} and~\eqref{eq: tov4}. Hence, we demand that these initial values should be the ones that render the contributions to the baryon mass at their respective orders, $\delta M_{\text{b,2}}$ and $\delta M_{\text{b,4}}$, zero. These constraints are

\begin{widetext}
\begin{subequations}
\begin{align}
    \label{eq: baryon2}
     \delta M_{\text{b},2} &= |\beta| \int_0^\infty \d r\,4\pi r^2\,\left(1-2\dfrac{\mu_0}{r}\right)^{-1/2} \left[\rt_{r,2} + \rt_{r,0}\left(\frac{\mu_2}{r-2\mu_0} - \dfrac{3}{2} \varphi^2\right)\right] = 0,\\
     \label{eq: baryon4}
    \delta M_{\text{b},4} &= |\beta|^2 \int_0^\infty \d r\,4\pi r^2\,\left(1-2\dfrac{\mu_0}{r}\right)^{-1/2} \Bigg[\rt_{r,4} + \rt_{r,2}\left(\frac{\mu_2}{r-2\mu_0} - \frac{3}{2}\varphi^2\right) \nonumber\\
    & \qquad\qquad\qquad\qquad\qquad\qquad\qquad\qquad + \rt_{r,0} \bigg( \frac{\mu_4}{r-2\mu_0} + \frac{3}{2} \frac{\mu_2^2}{(r-2\mu_0)^2}
    -\frac{3}{2}\frac{\mu_2}{r-2\mu_0}\varphi^2 + \frac{9}{8}\varphi^4 \bigg) \Bigg] =0.
    \end{align}
\end{subequations}
\end{widetext}

Just as in the usual integration of the fully nonlinear TOV system~\eqref{eq: tov full}, we need the EOS to close the system in each perturbative order \eqref{eq: gr tov}-\eqref{eq: tov4}. We simply use the HB EOS in the zeroth order, i.e., GR background, but higher-order equations include derivatives of the EOS relationship. In the $\mathcal{O}(\phi_\text{c}^2)$ order, from the Taylor series expansion of the EOS around the GR background, we obtain
\begin{equation}\label{eq: rt_2}
    \rt_2 = \left(\frac{\d\rt_0}{\d\pt_0}\right)_{\text{GR}}\pt_2 = \frac{\pt_2}{c_\text{s}^2},
\end{equation}
where $c_\text{s}$ is the speed of sound. In the next, $\mathcal{O}(\phi_\text{c}^4)$ order, the same expansion yields
\begin{equation}\label{eq: rt_4}
     \rt_4 =  \left ( \frac{\d\rt_0}{\d\pt_0}\right )_{\text{GR}}\pt_4 + \frac{1}{2}\left(\frac{\d^2\rt_0}{\d\pt_0^2}\right)_{\text{GR}}\pt_2^2.
\end{equation}

We adopted a piecewise polytropic EOS which has separate core and crust regions with different adiabatic indices, which are continuously connected at a transition density~\cite{Read:2009yp}. However, even though $\rt(\pt)$ is continuous for such an EOS, its derivatives are not at the transition pressure. As a result, the left and right derivatives in Eqs.~\eqref{eq: rt_2} and~\eqref{eq: rt_4} are discontinuous at the transition pressure. If one is not careful, this introduces numerical artifacts in the form of spikes if one performs finite differencing around the transition density.\footnote{Note that we could as well use a single polytrope or some other smooth EOS to avoid this problem, since the choice of EOS is not essential in developing and demonstrating our energy function framework. However, in order to facilitate comparison to past data~\cite{Unluturk:2025zie}, we decided to keep a piecewise polytropic EOS.}

We circumvent this numerical problem by detecting the artificial spike we would have if we tried to perform finite differencing at the transition density, and smoothing it out using a piecewise cubic Hermite interpolating polynomial as it preserves data monotonicity and shape, avoiding overshoots found in standard splines.

\section{Order of the phase transition\\at less negative $\beta$ values}
\label{appx: decreasing I_2}
%
\begin{figure*}
    \centering
    \includegraphics[width=\linewidth]{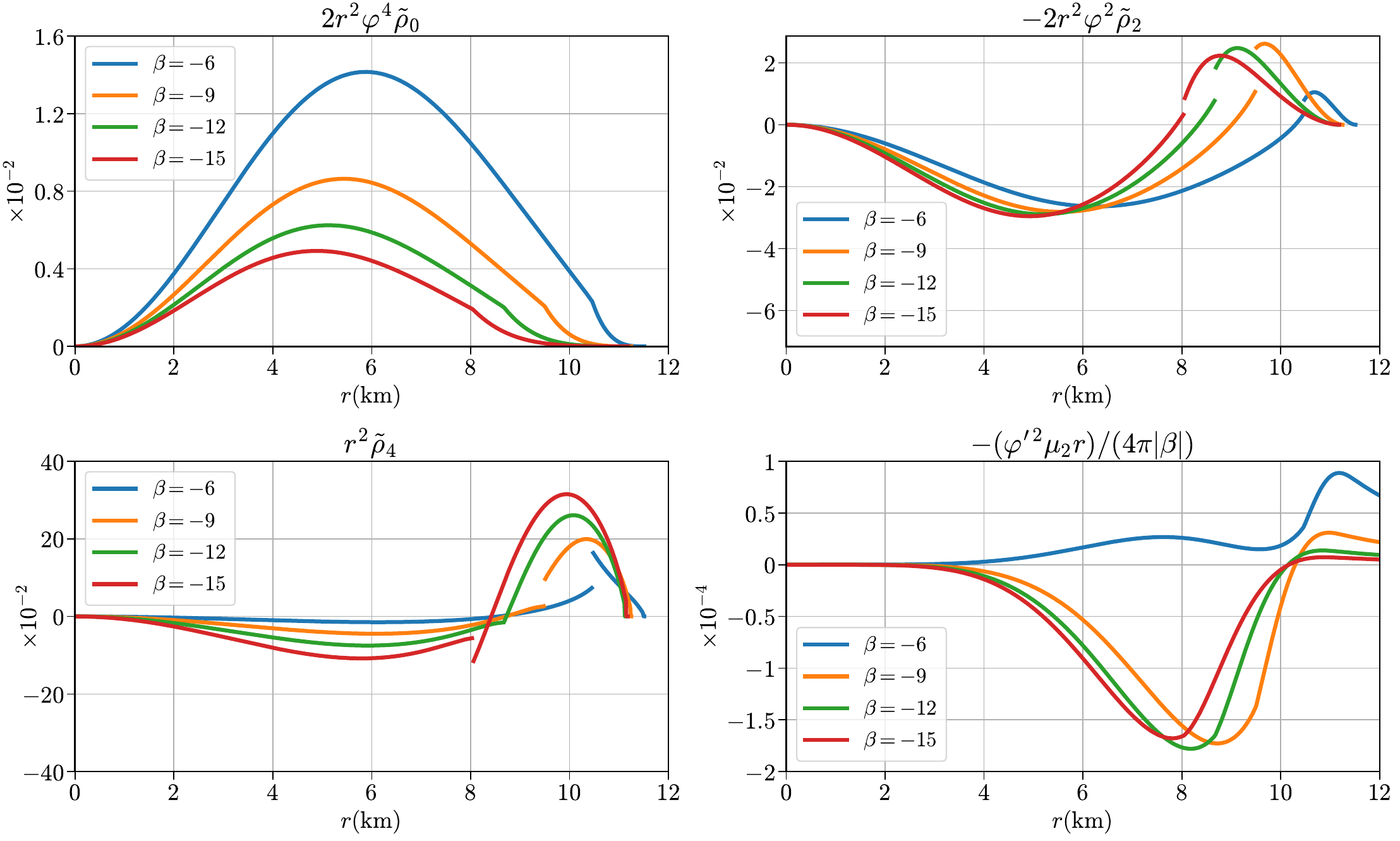}
    \caption{The same as Figure~\ref{fig: integral_terms}, but for less negative values of $\beta$. The top right panel demonstrates why $I_2$ decreases in magnitude instead of increasing in the $-\beta \lesssim 20$ region of Figure~\ref{fig: integral_terms}.}
    \label{fig: more integral_terms}
\end{figure*}

We explained in Sec.~\ref{sec: explaining first-order} that the main culprit for seeing first-order scalarization at very negative values of $\beta$ was the dominance of the negative $I_2$ term in Eq.~\eqref{eq: b integral} which grows in magnitude with $-\beta$. Note that this growth only happens when $\beta$ is sufficiently negative, $-\beta \gtrsim 20$ in Figure~\ref{fig: integral_terms}. Before this point, $I_2$ temporarily decreases in magnitude, though it still dominates to make $b_0$ negative. This is not a primary point for our purposes, since we mainly seek to understand why first-order scalarization becomes the norm for sufficiently negative $\beta$. Pinpointing the $\beta$ value where the transition switches from second- to first-order is not essential. Nevertheless, we can still explain the $-\beta \lesssim 20$ behavior by investigating how the bifurcation solution moves on the mass-radius curve as $\beta$ changes.

The decrease in the magnitude of the $I_2$ term for $-\beta \lesssim 20$ in Figure~\ref{fig: integral_terms} is due to how its integrand $- 2 r^2 \varphi^2 \rt_2$ behaves in the top right panel of Figure~\ref{fig: b integral terms}. Note that $- 2r^2  \varphi^2 \rt_2$ becomes positive near the stellar surface as explained in Sec.~\ref{sec: explaining first-order}. This positive region contributes less than the negative region for all $\beta$, and its contribution also grows more slowly than the negative region when $\beta$ becomes sufficiently negative. However, for less negative $\beta$, such as $-\beta \lesssim 20$ in Figure~\ref{fig: integral_terms}, the positive region of $- 2r^2  \varphi^2 \rt_2$ grows slightly faster than the negative one. This can be seen in Figure~\ref{fig: more integral_terms}, which is an analog of Figure~\ref{fig: b integral terms}, but we plot the integrands for different $-\beta \lesssim 20$ values. The transient relative growth of the positive region of $- 2r^2  \varphi^2 \rt_2$ decreases the magnitude of $I_2$, but never to the extent that $I_2$ becomes positive or subdominant to other terms in $b_0$ in Eq.~\eqref{eq: b integral}.

We strongly suspect this behavior for $-\beta \lesssim 20$ to be related to the mass-radius curves for our specific choices of EOS, which can be seen in Figure~\ref{fig: mass radius}. The ``usual'' behavior on this curve is that increasing ADM mass moves us to lower stellar radii. This is one reason that less massive stars are usually also less compact and less stiff. However, note that there is a small ``exceptional'' region in each mass-radius curve in Figure~\ref{fig: mass radius} where the opposite behavior occurs: the less massive stars are smaller in radius. Compactness still monotonically decreases with decreasing mass, yet, the slope of the compactness-mass curve is considerably less steep in this exceptional region on the mass-radius curve. Recall that we used the dependence of the bifurcation solution and its stiffness on $\beta$ in our explanations in Sec.~\ref{sec: explaining first-order}, hence it is not totally surprising that this special region deviates from the norm in some aspects.

To be exact, note that the bifurcation solutions for $-\beta \lesssim 20$ more or less coincide with the above region of the mass-radius curve we described. Hence, the less usual dependence of the stellar radius on the stellar mass inflates the importance of the positive region of $- 2 r^2 \varphi^2 \rt_2$ in the top right panel of Figure~\ref{fig: more integral_terms}, which leads to the temporary decrease in $I_2$ as $\beta$ becomes more negative, until $\beta \sim -20$. 

We emphasize that $I_2$ becomes negative well before $\beta \sim -20$, so the dominance of $I_2$ and the switch to first-order scalarization is possible despite a decreasing $I_2$ in magnitude. Moreover, more negative $\beta$ is destined to bring the bifurcation solution to lower ADM masses, where the mass-radius dependence is the more common one, and our explanations in Sec.~\ref{sec: explaining first-order} work without caveats. Nevertheless, exploring the phase transition properties at less negative $\beta$ values provides a more thorough understanding of scalarization. Since the exceptional region of the mass-radius curve depends on EOS, this also suggests that the scalarization phase transition might behave in unusual ways for very exotic EOS.

\end{document}